\documentclass[twocolappendix,numberedappendix]{emulateapj}
\usepackage{natbib}
\usepackage{apjfonts}
\usepackage{placeins}
\usepackage{multirow}
\usepackage{amsmath}
\usepackage{amssymb}
\usepackage{epsfig}
\usepackage{graphicx}
\usepackage{color}
\newcommand{\kms}{\mbox{km s$^{-1}$}}

\newcommand{\HI}{\ion{H}{1}}
\newcommand{\hii}{\ion{H}{2}}
\newcommand{\sii}{\ion{S}{2}}

\newcommand{\co}{\mbox{CO$(1\rightarrow0)$}}
\newcommand{\hcn}{\mbox{HCN}}
\newcommand{\sio}{\mbox{SiO}}

\newcommand{\hcop}{\mbox{HCO$^{+}$}}
\newcommand{\ntwohp}{\mbox{N$_\mathrm{2}$H$^{+}$}}

\newcommand{\cother}{\mbox{$^{13}$CO$(1\rightarrow0)$}}

\newcommand{\degrees}{\arcdeg}

\shorttitle{Effect of Feedback in the Carina Nebula}
\shortauthors{Rebolledo et al.}

\begin{document}
\submitted{Accepted for publication in the Astrophysical Journal on 29th of December 2020}

\title{The Carina Nebula and Gum 31 molecular complex III:  The distribution of the 1-3 GH\MakeLowercase{z} radio continuum across the whole nebula.}

\author{David Rebolledo$^{1,2}$, Anne J. Green$^{3}$, Michael G. Burton$^{4,5}$, Shari L.\ Breen$^{6}$, Guido Garay$^{7}$}

\affil{$^{1}$Joint ALMA Observatory, Alonso de C\'ordova 3107, Vitacura, Santiago, Chile; david.rebolledo@alma.cl\\
$^{2}$National Radio Astronomy Observatory, 520 Edgemont Road, Charlottesville, VA 22903, USA\\
$^{3}$Sydney Institute for Astronomy, School of Physics, The University of Sydney, NSW 2006, Australia\\
$^{4}$Armagh Observatory and Planetarium, College Hill, Armagh, BT61 9DG, Northern Ireland, UK\\
$^{5}$School of Physics, The University of New South Wales, Sydney, NSW, 2052, Australia\\
$^{6}$SKA Organisation, Jodrell Bank Observatory, SK11 9DL, UK\\
$^{7}$Departamento de Astronom\'ia, Universidad de Chile, Camino el Observatorio 1515, Las Condes, Santiago, Chile
}

\begin{abstract}

We report the most detailed $1-3$ GHz radio continuum emission map of the nearest region of massive star formation, the Carina Nebula. As part of a large program with the Australia Telescope Compact Array, we have covered $\sim$ 12 $\deg^2$, achieving an angular resolution of  $\sim 16\arcsec$, representing the largest and most complete map of the radio continuum to date.  Our continuum map shows a spectacular and complex distribution of emission across the nebula, with multiple structures such as filaments, shells, and fronts across a wide range of size scales. The ionization fronts have advanced far into the southern and northern region of the Galactic Plane, as fronts are clearly detected at distances $\sim$ 80 pc from the stellar clusters in the center. We estimated an ionization photon luminosity $Q_\mathrm{H}=(7.8 \pm 0.8) \times 10^{50}$ s$^{-1}$ which corresponds to $\sim 85\%$  of the total value obtained from stellar population studies. Thus, approximately $15\%$ of the ionizing flux has escaped from the nebula into the diffuse Galactic Interstellar Medium. Comparison between radio continuum and the hydrogen atomic and molecular gas maps offers a clear view of the bipolar outflow driven by the energy released by the massive stellar clusters that also affects the fraction of molecular gas across the nebula. Comparison between 8$\mu$m and 70$\mu$m emission maps and the radio continuum reveals how the hot gas permeates through the molecular cloud, shapes the material into features such as pillars, small shells and arc-like structures, and ultimately, escapes.

\end{abstract}


\section{INTRODUCTION}

In this paper we release the third data set produced by our long term multi-wavelength and multi-scale observing campaign to study the different gas phases of the Carina Nebula Complex (CNC) and its close neighbor, the \hii\ region Gum 31.  The CNC-Gum 31 complex has in total more than 65 O-stars and multiple new protostars (\citealt{2006MNRAS.367..763S}; \citealt{2010MNRAS.406..952S}), representing our nearest analogue of extreme star forming regions such as 30 Doradus in the Large Magellanic Cloud.  Because Carina is relatively nearby (2.3 kpc, \citealt{2008hsf2.book..138S}), we can resolve a wide range of size structures, from the entire distribution of the gas across the complex to the dense cores in localized regions (\citealt{2019MNRAS.483.1437C}; \citealt{2020ApJ...891..113R}).  Optical and infrared observations have provided ample evidence for active star formation in the dust pillars, with more than 900 young stellar objects identified (\citealt{2010MNRAS.405.1153S}). 

In the first paper of the campaign (\citealt{2016MNRAS.456.2406R}, hereafter Paper I), the molecular gas component of the CNC-Gum 31 molecular complex was obtained with the Mopra telescope. Using the $\co$ and $\cother$ maps, we estimated the molecular gas column density distribution across the complex.  In parallel, the total gas column density and dust temperatures were estimated from Herschel maps (Hi-GAL, \citealt{2010PASP..122..314M}).  The strong impact of the massive star clusters located at the center of the Nebula was printed in the significant dust temperature variations across the complex.  Detailed comparison between the total gas column density derived from dust emission maps, and the molecular column density derived from CO maps allowed us to chart the variation of the CO-to-$N_\mathrm{H2}$ conversion factor across the CNC, and link this variation to the differences in gas temperature and level of stellar feedback.

The second paper of the project published high sensitivity and high resolution maps of the \HI\ 21-cm line towards the CNC obtained with the Australia Telescope Compact Array (\citealt{2017MNRAS.472.1685R}, hereafter Paper II).  This detailed map of the atomic gas revealed a complex filamentary structure across the velocity range that crosses the Galactic disk.  Absorption features detected in the high resolution \HI\ spectra due to the presence of compact and diffuse continuum sources, allowed us to identify the cold component of the atomic gas, and in some particular cases, the line optical depth and the spin temperature, two quantities difficult to obtain from pure line emission maps.  Detection of \HI\ self-absorption also revealed the presence of cold neutral gas in some regions, signaling the places where the important phase transition between atomic and molecular gas is occurring.

In order to investigate the effect of stellar feedback at smaller scales, we carried out Mopra observations toward 60 massive clumps in the CNC to characterize their physical and chemical evolution (\citealt{2019MNRAS.483.1437C}).  This study showed that the clumps in Carina are warmer, less massive, and show less emission from the four most commonly detected molecules, \hcop, \ntwohp, \hcn, and HNC, compared to clumps associated with masers in the Galactic Plane (\citealt{2009MNRAS.392..783G}; \citealt{2012MNRAS.420.3108G}; \citealt{2016MNRAS.461..136A}). This result provided support for the scenario in which the high radiation field of nearby massive stars is dramatically affecting its local environment, and therefore the chemical composition of the dense clumps.

From the sample of massive clumps observed with Mopra, we undertook ALMA follow up observations towards two selected regions with very different physical properties.  One region was located at the center of the nebula and multiply affected by the stellar feedback from high-mass stars, while the other region was located farther south and is less disturbed by the massive-star clusters.  Our study revealed that the region at the center of the nebula was forming less but more massive cores than the region located in the south, suggesting that the level of stellar feedback effectively influences the fragmentation process into clumps (\citealt{2020ApJ...891..113R}).  Jeans analysis indicates that the observed core masses in the region less affected by the massive stars are consistent with thermal fragmentation, but turbulent Jeans fragmentation might explain the high masses of the cores identified in the region in the center of Carina. Separate analysis of the detected molecular lines and the dust column density provided consistent evidence for a higher level of turbulence in the gas more affected by the stellar feedback.
 
In this work, we present high resolution observations of the $1-3$ GHz radio continuum that trace the ionized component of the Interstellar Medium (ISM) in the CNC-Gum 31 region.  The paper is organized as follows:  In Section \ref{obs} we detail the main characteristics of the observations and the procedure to calibrate and image the data.  In Section \ref{results} we show the complex structure of the ionized gas, and discusses the features revealed by the map.  Section \ref{discuss} compares the radio continuum map with other gas tracers, and shows detailed maps towards a few interesting region identified in the radio continuum map. Section \ref{summary} summarizes our main conclusion.

\section{DATA}\label{obs}

\subsection{Observations}   
The radio continuum observations reported in this work are part of the large program project conducted with the ATCA synthesis imaging telescope to map the CNC-Gum 31 region.  In Paper II we reported the part of the survey that covers the \HI\ 21-cm line data, and in this paper we will focus on the details of the radio continuum data.  

\begin{table}
\caption{Spectral setup of ATCA observations.}
\centering
\begin{tabular}{ccc}
\hline\hline
\multicolumn{3}{c}{Continuum} \\
\hline
Width (MHz) & Cent. Freq. (MHz) & Windows    \\
\hline
2048 & 2100  & 2    \\ 
\end{tabular}

\begin{tabular}{cccc}
\hline\hline
\multicolumn{4}{c}{Lines} \\
\hline
Zoom Band & Cent. Freq.   &  Num. zooms & Line covered   \\
& (MHz) & & \\
\hline
z1 & 1420 & 4 &   \HI\ 21 cm     \\ 
z2 & 1651 & 6 & H158$\alpha$ recombination  \\
z3 & 1720.75 & 4 & 1720 MHz OH maser  \\
z4 & 1666.25  & 6 & 1665/1667 MHz OH masers    \\ 
\hline
\end{tabular}
\label{spectral-setup}
\end{table}

\subsubsection{Spectral Setup}   
The Compact Array Broadband Backend (CABB, \citealt{2011MNRAS.416..832W}) was used in the CFB 1M-0.5k configuration. Table \ref{spectral-setup} shows the spectral setup of our observations. This spectral configuration allowed us to cover the continuum with two redundant wide windows, and use up to 16 zoom bands to cover the targeted lines.  All the 4 polarization products are computed. Both continuum windows have 2048$\times$ $1-$MHz channels, producing a frequency coverage from 1.076 GHz to 3.124 GHz.  The continuum windows redundancy was implemented in order to minimize the impact of correlator problems that can affect significant parts of the spectral coverage.

\subsubsection{Spatial Setup and $u$-$v$ coverage}   
CNC region features a complex emission distribution ranging from arc-second to the degree scale, in addition to a strong $\approx$2 Jy radio emission associated with the luminous star $\eta$ Car. In order to minimize telescope artifacts and maximize the dynamic range of the image, we selected an optimal set of array configurations, namely, 6A, 6B, 6C, 1.5A, 1.5B, 1.5D, 750A, 750B, 750C, 750D, and EW352.  With this selection, we are sampling $u$-$v$ baselines from 30.6 m to 5 km.  From 30.6 m to 795.9 m, we achieved an almost fully uniform sampling in intervals of 15.3 m. Duplication of shorter baselines was necessary as radio frequency interference often affected the lower part of the frequency band, and was strongest on the shorter baselines.  Substantial removal of corrupted samples was required. Hence, to preserve our sensitivity for the largest scale structures, observations with the EW352 array were undertaken. Each mosaic in a given array configuration was observed for an average of 12 hours.

\begin{figure}[tbph]
\centering
\begin{tabular}{c}
\epsfig{file=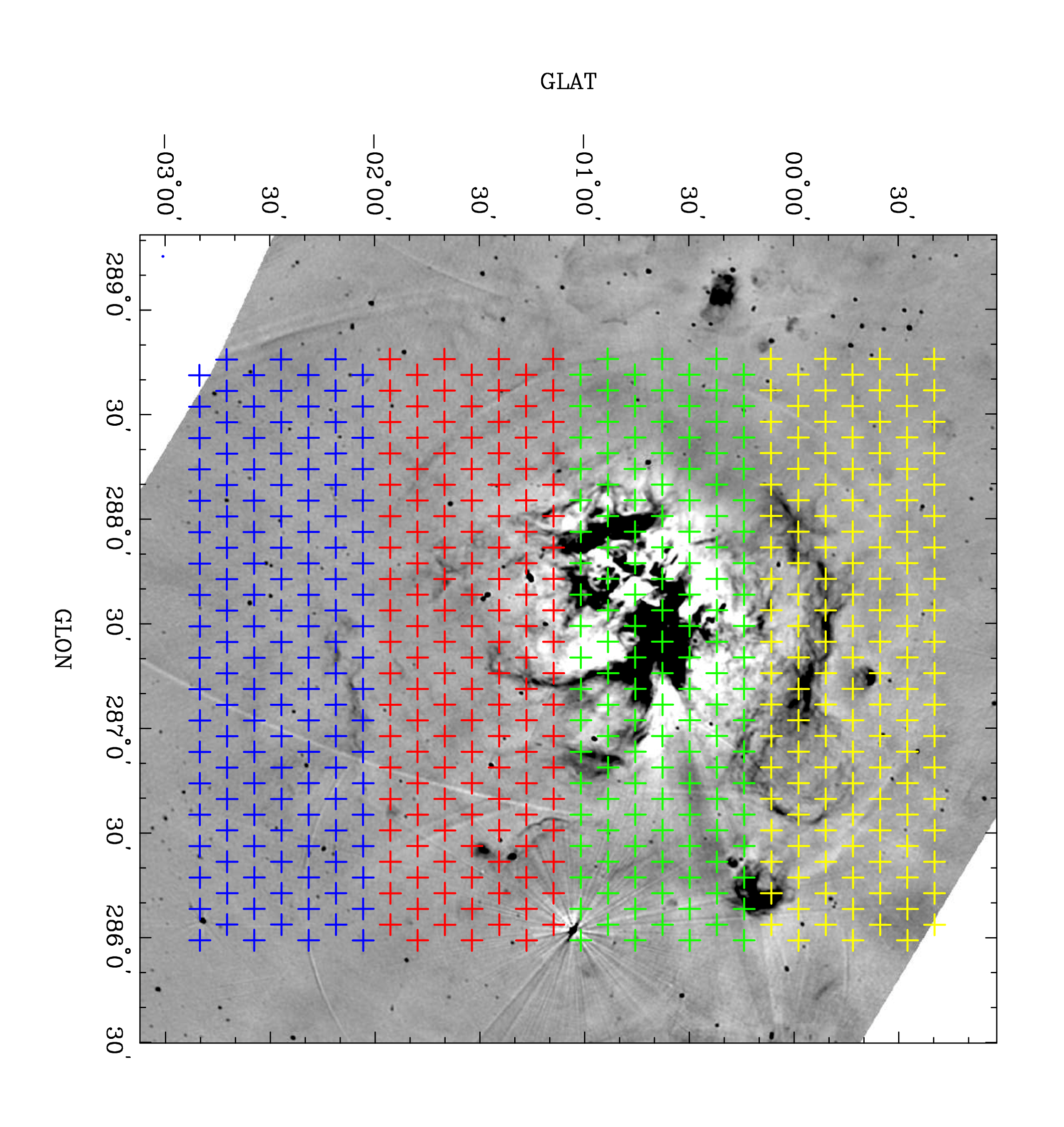,width=0.95\linewidth,angle=90}
\end{tabular}
\caption{Mosaic pattern of our ATCA observations overlaid in a map of the 0.835 GHz continuum emission in the CNC-Gum 31 complex obtained with the Molonglo Observatory Synthesis Telescope (\citealt{2014PASA...31...42G}).  In total, 532 pointings were observed, divided into 4 independent mosaics of 133 pointings each (illustrated with a different color).  Each mosaic was observed using 11 array configurations to maximize $u-v$ coverage.  Thus, our observing campaign completed 44 sessions with the ATCA.  Each array configuration observation was individually calibrated before combination for final imaging.}
\label{cont_most}
\end{figure}

\begin{figure*}[tbph]
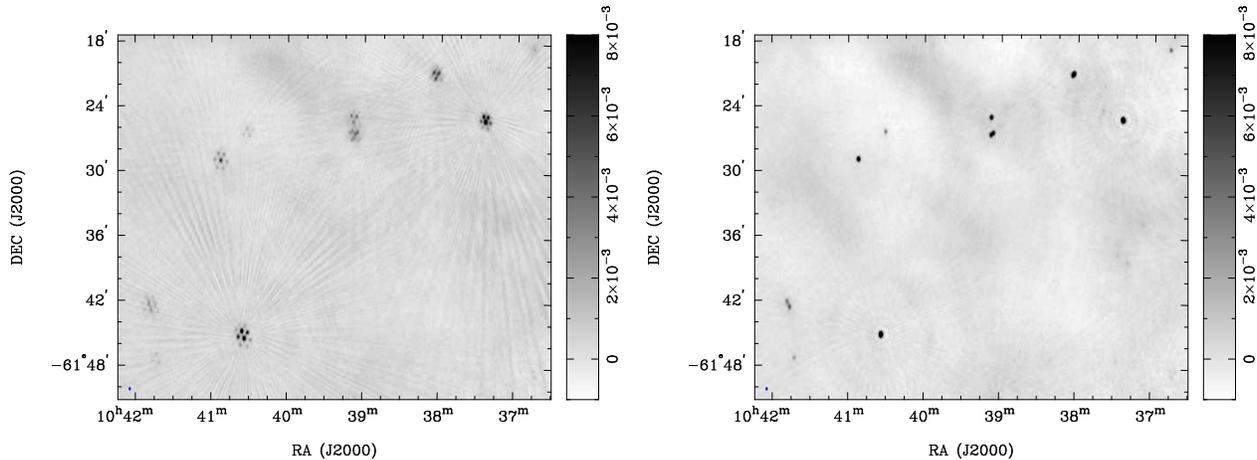

\centering
\begin{tabular}{cc}
\epsfig{file=Figure2_a.ps,width=0.45\linewidth,angle=0} & \epsfig{file=Figure2_b.ps,width=0.45\linewidth,angle=0} 
\end{tabular}
\caption{Left:  Image of a group of point sources affected by the residual position offsets introduced after we directly correct the phase center in each field in the visibility domain.  A point source appears as a cluster of sources, each of them belonging to different mosaic fields that cover the region.  Right: Same region in the left panel, but after our correction approach has been applied to the data (see Figure \ref{figure_approach}).  Now, unresolved sources are correctly displayed.}
\label{figure_phasewrong}
\end{figure*}

Figure \ref{cont_most} shows a map of the 0.835 GHz continuum emission in the CNC-Gum 31 complex obtained with the Molonglo Observatory Synthesis Telescope (\citealt{2007MNRAS.382..382M}; \citealt{2014PASA...31...42G}).  This figure also shows the mosaic pattern we used to cover the targeted area.  Our observations cover the region $285.8\degrees \lesssim l \lesssim 289\degrees$ and $-3.0\degrees \lesssim b \lesssim 1.0\degrees$. In total, we observed 532 pointings, which were divided into four mosaics each one composed of 133 pointings.  The center of each pointing was chosen to ensure Nyquist sampling at the highest observed frequency.  The data were taken between April of 2011 and February of 2015 using 11 array configurations for each mosaic, totaling 44 complete observing runs.  Although the resulting $u$-$v$ sampling of an individual pointing may vary across the observed field, our observational strategy provided a nearly uniform $u$-$v$ coverage across the mosaic.

\subsection{Calibration}
Flagging and calibration of the data was performed using the standard techniques with the Multichannel Image Reconstruction, Image Analysis and Display (MIRIAD)  data reduction package (\citealt{1995ASPC...77..433S}).  The primary flux calibrator was PKS B1934--638, which was observed at least once per observing run.  This source was used for passband and absolute flux calibration, with an assumed flux density of 14.86 Jy at 1.420 GHz.  The phase calibrator was PKS B1036--697, which was observed for 3 minutes interleaved in each mosaic cycle observations, which typically took $\sim$ 20 minutes.

\subsection{Correction of incorrect phase center due to CABB software limitation} 
Until May of 2012, the CABB system was affected by a software limitation during observations.  Before this date, the system was able to only read the first eight characters of the name of each mosaic field when setting the phase center position during observations.  Our observing campaign started in 2011, and our mosaic field names were longer than eight characters.  Fortunately, this software limitation did not affect the antennas, so they were pointing correctly in the sky during observations.  The resulting effect of this software limitation was that, although the coordinate center information was correctly written in the header of the visibilities,  the recorded $u-v$ data were computed using the coordinate center of another field.  Approximately, 50\% of our data were affected by this problem, therefore appropriate phase center correction was needed before final imaging.  This correction can be performed following different methods depending on the characteristic of the data and the requirements of the imaging process.  

In our particular case, we want to combine all the observed array configurations before imaging, so the correction must be done in the visibility domain.  This approach consists in editing the header in the corrupted visibility data to have the coordinates actually used to compute the $u-v$ data during observations.  Then, one must change the coordinates of the phase center in the visibility data to the correct values, and recalculate the visibilities using this new reference position.  Although this method is a direct and fast approach, it is not perfect.  Some small residual position offsets are detected once the corrected visibilities are imaged.  These offsets are not significantly large, but they might represent an issue for images with high spatial resolutions.  Figure \ref{figure_phasewrong} (left) shows an example of the residual offset effect in point sources present in some of our fields.  An unresolved source appears as a cluster of sources, each of them belonging to the visibilities of the mosaic field neighbors.  The position offsets are larger for fields located farther away from the reference center position used to calculate the visibilities during observations.

\begin{figure*}[tbph]
\centering
\epsfig{file=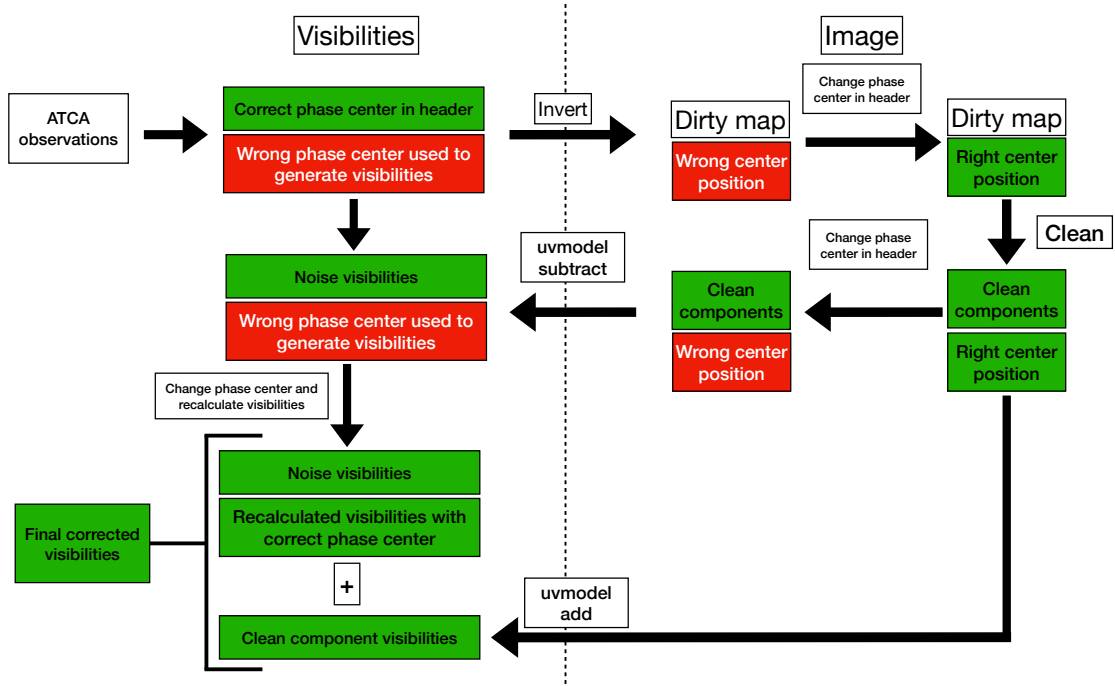,width=0.7\linewidth,angle=90}
\caption{Schematic description of the approach used to correct the wrong phase center in the data taken before May 2012.  The vertical dotted line separates the visibility (left) and the image (right) domains.  Our approach starts by correcting the phase center in the image domain to generate the image clean components.  Then, we subtract the clean components from the corrupted visibilities, and add them back after the emission-free visibilities are corrected in the visibility domain.  This method avoids the position offset effect illustrated in Figure \ref{figure_phasewrong}.}
\label{figure_approach}
\end{figure*}

In order to properly correct the phase center in the visibilities, an alternative method had to be developed.  We implemented an approach that combines different steps in the image and in the visibility domain.  A schematic diagram is presented in Figure \ref{figure_approach}, and here we review the main steps:

\begin{enumerate}

\item The first step was to generate the dirty images for each field from the corrupted visibilities.  Because these uv-data were computed using a wrong phase center, then the images have a wrong position in the sky.  We correct the position of the field image by manually editing the coordinates of the center in the header.  The corrected dirty images have the correct position in the sky.

\item The second step consisted in cleaning the dirty map generated in the previous step to obtain the clean components.  These are converted into visibilities using the {\it uvmodel} task in MIRIAD, and saved.

\item In the third step, we edited the header of the clean components to put them back  in the wrong center position as they were in step 1.  Thus, we could subtract them from the corrupted visibilities to generate emission-free uv-data.  Then, we applied the phase center correction in the subtracted visibilities.

\item Finally, the phase corrected emission-free uv-data generated in step 3 were combined with the visibilities created from the clean components generated in step 2.

\end{enumerate}

Figure \ref{figure_phasewrong} (right) shows the resulting image after our approach to correct the phase center is applied.  Now, we clearly recover the point sources, and the residual position offsets have disappeared.  After this method is applied to the visibilities of each field, we combine all the $u-v$ data to continue with the imaging process.

\subsection{Imaging}\label{imaging}
In order to optimize uv-coverage, but at the same time to minimize spectral index effects in the imaging, we have split our 2 GHz frequency coverage into two 1 GHz bands ($1-2$ GHz and $2-3$ GHz) to generate two separate maps. Imaging was performed using a joint deconvolution approach with the MIRIAD task {\it invert}  (\citealt{1996A&AS..120..375S}).  A dirty image for each individual pointing is generated using a standard grid and Fast Fourier Transform (FFT).  All the dirty images for each mosaic were then combined linearly to create a mosaicked dirty map.  We have selected a robust parameter equal to 0.5 to generate the dirty images in both maps, the $1-2$ GHz and $2-3$ GHz frequency ranges.

\begin{figure*}[tbph]
\centering
\epsfig{file=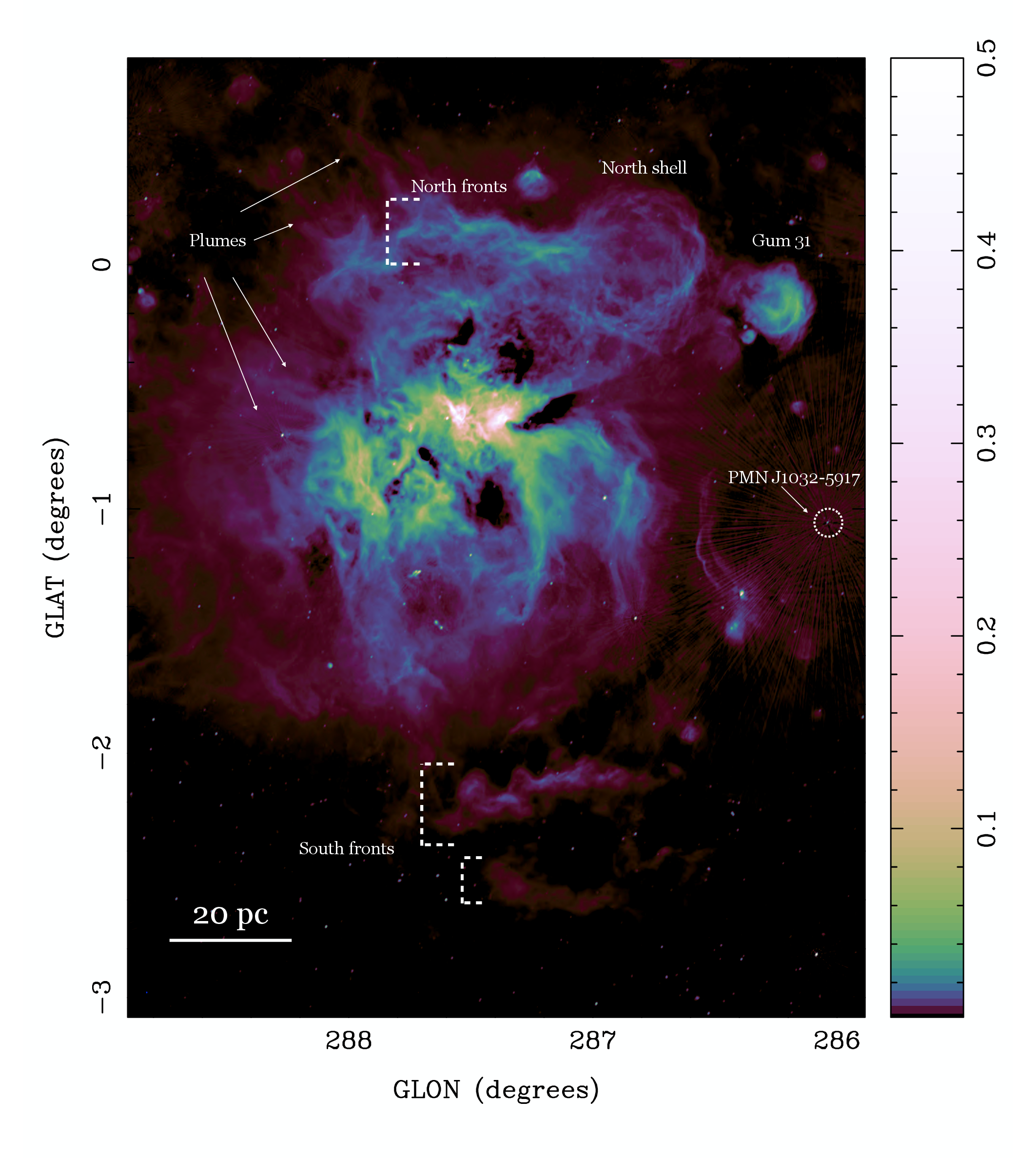,width=0.9\linewidth,angle=0}
\caption{ATCA $1-3$ GHz continuum image of the CNC-Gum31 complex. The image size is 3$\times$4 deg$^2$, which was obtained with 532 pointing grid observations. Color bar is in Jy/beam. Our map provides an unprecedented high resolution and high fidelity radio continuum image of the Carina Nebula and surrounding regions.  The spectacular and complex structure of the emission is revealed, with features similar to shells, filaments and ionization fronts.  A large population of compact sources are also clearly detected across the surveyed area, the majority of which are extragalactic radio sources.  Dashed white circle shows the position of the strong radio source PMN J1032-5917, which was subtracted from the uv-data.}
\label{carina_conti_full}
\end{figure*}

Self-calibration can be utilized to improve the image quality in regions with strong point sources.  The effectiveness of self-calibration to improve a particular image depends on several factors such as the flux density of the source, and the complexity of the emission in the field.  In our case, we have performed self-calibration for one particular object, the radio source PMN J1032-5917.  This radio source has a flux density $\sim$ 2.8 Jy at 1.4 GHz, and it is the strongest source in the field.   Because this source can be detected by multiple pointings, then self-calibration can be challenging, as the strong source will appear at the edge of the primary beam of more distant fields.  Our strategy applied self-calibration steps independently on each field surrounding the strong source.  We use phase corrections only, because amplitude corrections in general are unstable and sometimes they introduced artifacts in the image.  Once self-calibrated, the radio source PMN J1032-5917 was subtracted from the uv-data used to generate the final images.

As in Paper II, we used the Maximum Entropy algorithm (\citealt{1999ASPC..180..151C}) for deconvolution.  MIRIAD task {\it mosmem} was used to generate the models, as it works better in recovering the diffuse emission compared to the CLEAN algorithm.  Another advantage of {\it mosmem} is that this method offers the capability of combining interferometric data and single dish observations in the deconvolution process. This combination is achieved by giving a single dish map as the initial solution during deconvolution.  The algorithm then forces the deconvolution towards values in the single dish image for the spatial frequencies without interferometric data.  For the lower half frequency range of the 2 GHz band ($1-2$ GHz), we have used the Parkes single dish image from the $\chi$-pass survey (\citealt{2014PASA...31....7C}).  This survey used the Parkes radio telescope to map the 1.4 GHz continuum emission of the sky south of declination +25$\degrees$ at high sensitivity (40 mK) and spatial resolution (14.4\arcmin).  For the upper half frequency range ($2-3$ GHz), the single dish map was taken from the S-band Polarisation All Sky Survey (S-PASS) survey (\citealt{2019MNRAS.489.2330C}). S-PASS mapped the polarized emission at 2.3 GHz of the southern sky at declinations < -1 $\degrees$ with the Parkes radio telescope.  Stokes {\it I} maps have a mean sensitivity of 9 mK, and a beam resolution of 8.9\arcmin.  

The two continuum images created from the two 1 GHz bands (at $\sim$ 1.4 GHz, and $\sim$ 2.4 GHz) were then averaged to make the final continuum image.  In Figure \ref{carina_conti_full} we show the resulting image.  The beam resolution of the image is 24\farcs4 $\times$ 15\farcs8, and the rms sensitivity is 2 mJy/beam.  Both radio continuum images (at $\sim$ 1.4 GHz, and $\sim$ 2.4 GHz) and the averaged final image will be available to the community in the website of the project\footnote{http://www.physics.usyd.edu.au/sifa/carparcs/public/data.php}.

\section{Results}\label{results}

\subsection{Morphology of the radio continuum emission}\label{morphology}

Figure \ref{carina_conti_full} reveals a spectacular view of the $1-3$ GHz radio continuum emission in the CNC-Gum 31 complex.  A complex distribution of emission at multiple scales is clearly recovered, with many features distinguishable across the region such as shells and filaments.  Emission is easily detected across the whole field, spanning $\sim$ 140 pc from north to south of the Galactic Plane.  The region that encloses the brightest emission is located in the center of the nebula at ($l$,$b$) $\sim (287.5,-0.6)$, where the massive star clusters are spatially distributed.  Most of the bright emission ($\gtrsim$ 0.02 Jy/beam) are located in a radius of $\sim$ 20 pc around this center.  

Massive star clusters had ionized the ambient media in the CNC, and disrupted the molecular cloud that gave birth to the first generation of stars.  Several star clusters are located here, such as Trumpler 16 ($\eta$ Car is a member), Trumpler 14, Trumpler 15, Bochum 10 and Bochum 11.  In total, 65 O stars, 57 B stars, and 3 Wolf-Rayet stars are present in the region, producing an ionizing photon luminosity of $Q_\mathrm{H}\simeq 9.1 \times 10^{50}$ s$^{-1}$, and a total mechanical luminosity of stellar winds of $L \simeq 10^5\ L_\odot$ (\citealt{2006MNRAS.367..763S}).  By using the integrated emission from a low resolution Parkes 3.4 cm radio continuum map, \citet{2007MNRAS.379.1279S} estimated that the hydrogen ionizing photons absorbed by the gas was $\sim 6.9 \times 10^{50}$ s$^{-1}$.  Therefore, about 25\% of the ionizing flux produced by the stars has been able to escape from Carina.  Following the same approach detailed in \citet{1990ApJ...354..165S}, we estimate $Q_\mathrm{H}$ from the radio continuum emission by using

\begin{equation}
\begin{aligned}
Q_\mathrm{H} = 5.59\times 10^{48} \frac{1}{1+f_i \left<\mathrm{He}^{+}/(\mathrm{H}^{+}+\mathrm{He})\right>} \\ 
			 \times T_e^{-0.45} \left(\frac{\nu}{\nu_5}\right)^{0.1} S_{\nu}D^{2}  \mathrm{s}^{-1},
\end{aligned}
\end{equation}

\noindent where $f_i$ is the fraction of He recombination photons that can ionize H, $T_e$ is the electron temperature, $\nu$ is the frequency in GHz ($\nu_5 = 5.0$ GHz), $S_\nu$ is the flux, and D is the distance.  As in \citet{2007MNRAS.379.1279S}, we assumed that $f_i=0.65$, the He/H abundance ratio is 0.1, and that the volume of the He$^+$ region is half of the H$^+$ region.  To measure the integrated radio flux, we have summed all the emission inside the rectangle shown in Figure \ref{multiphase_hi_cont}. This box is approximately the same size as the box used by \citet{2007MNRAS.379.1279S} to estimate the flux.  In this case, we have used the radio continuum map created from the upper frequency range ($2-3$ GHz), as this range is much less affected by RFI.  Our goal is to minimize the impact in the total flux of the missing visibilities. The flux uncertainty has been estimated assuming a conservative 10\% flux error, which is typical for the ATCA 16 cm band.  Inside this box, a $S_{\nu} =(1.85 \pm 0.19)\times 10^3$ Jy was obtained, which corresponds to a value of $Q_\mathrm{H}=(7.8 \pm 0.8) \times 10^{50}$ s$^{-1}$.  This value is higher than the value obtained by \citet{2007MNRAS.379.1279S} from the 3.4 cm radio continuum map, but still smaller than the value obtained by \citet{2006MNRAS.367..763S} from the stellar population.  According to our estimation of Q$_\mathrm{H}$, between $(15 \pm 8)$\% of the ionizing flux was able to leak out of the nebula.  The difference between our value and the value estimated previously for Q$_\mathrm{H}$ might be related to the different sizes used to define the box to calculate the integrated flux, and also due to differences in the sensitivity of the different radio maps.

\subsubsection{Center of the CNC}\label{center_cnc}
\begin{figure*}[tbph]
\centering
\begin{tabular}{c}
\epsfig{file=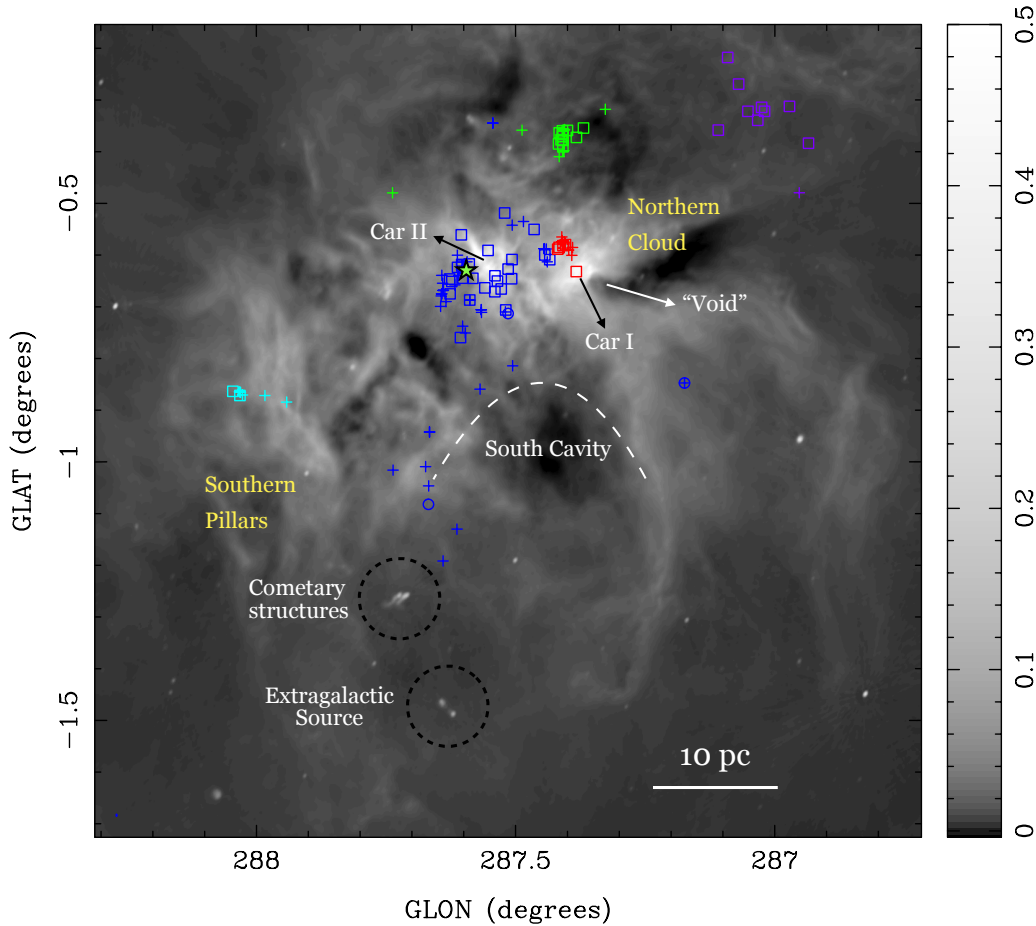,width=0.75\linewidth,angle=90}
\end{tabular}
\caption{View of the radio continuum emission at $1-3$ GHz in the central region of the CNC.  Symbols show the position of the massive stars, with different colors illustrating different star clusters.  Crosses show  O stars, squares show B stars, and open circles show Wolf Rayet stars.  Blue symbols show Trumpler 16 and Collinder 228 stellar clusters, red symbols show Trumpler 14, green show Trumpler 15, violet show Bochum 10 and cyan show Bochum 11 cluster as listed in \citet{2006MNRAS.367..763S}.  Black arrows show the position of the the two \hii\ regions analyzed by \citet{2001MNRAS.327...46B}, Car \ion{}{1} and Car \ion{}{2}.  White arrow shows the ``void'' region where radio continuum emission seems to be absent.  The curved white dashed line delineates the southern cavity discussed in the text.  The black dashed circles shows the cometary objects and the extragalactic source discussed in Section \ref{particular}.}
\label{carina_conti_cent}
\end{figure*}

Figure \ref{carina_conti_cent} provides an unprecedented detailed view of the heart of the CNC along with the positions of massive stars in the region.  $\eta$ Car has a radio emission peak of 0.8 Jy/beam, and represents the strongest source in the field.  This massive star is a member of Trumpler 16, which along with Trumpler 14 have shaped the region, producing two bright \hii\ regions, Car \ion{}{1} and Car \ion{}{2} (\citealt{2001MNRAS.327...46B}).  Both regions have peak emission $\sim 0.5$ Jy/beam, and features consistent with both regions being influenced by nearby stars.  \citet{2001MNRAS.327...46B} made detailed ATCA observations of the 4.8 GHz continuum emission and the H110$\alpha$ recombination line towards Car \ion{}{1} and Car \ion{}{2}.  The velocity information provided by the recombination line data revealed that Car \ion{}{1} and Car \ion{}{2} are expanding ionization fronts from Trumpler 14 and 16 respectively.  Our much larger image allows us to extend this detailed view towards the surrounding medium.

In the east, where the Southern Pillars are located (Figure \ref{cont_co}), several ionizing fronts are clearly identified.  As it will be shown in Section \ref{multi-phase}, these fronts are located at different levels along the surface of the colder gas, interleaved between the pillars that point to the massive clusters Trumpler 14 and 16.  The fronts run perpendicular to the direction vector to Trumpler 16, so it is clear this massive cluster dominates the influence in the gas in the eastern part of the nebula.  

Towards the west, the radio emission distribution is consistent with a strong influence from Trumpler 14.  A bifurcation of the emission is clearly seen starting at the position of Car \ion{}{1}. One arm extends to the south of the Galactic Plane forming the eastern wall that encloses the south cavity (see below), and the other arm extends beyond the location of the Northern Cloud to the west (Figure \ref{cont_co}).  Above this bifurcation feature, there is a region where almost no radio continuum has been detected.  The shape of this ``void'' is elongated towards the west, and it begins in the arc-like structure identified in Car \ion{}{1}.  We speculate that dense gas besides Car \ion{}{1} is blocking the expansion of the ionization fronts from Trumpler 14 farther to the west, producing this arc-like structure identified in the radio.  These arc-like structures in Car \ion{}{1} were also reported in \citet{2001MNRAS.327...46B}, and more recently, in high resolution observations with ALMA in \citet{2020ApJ...891..113R} (see Section \ref{NC}). 

\begin{figure*}[tbph]
\centering
\epsfig{file=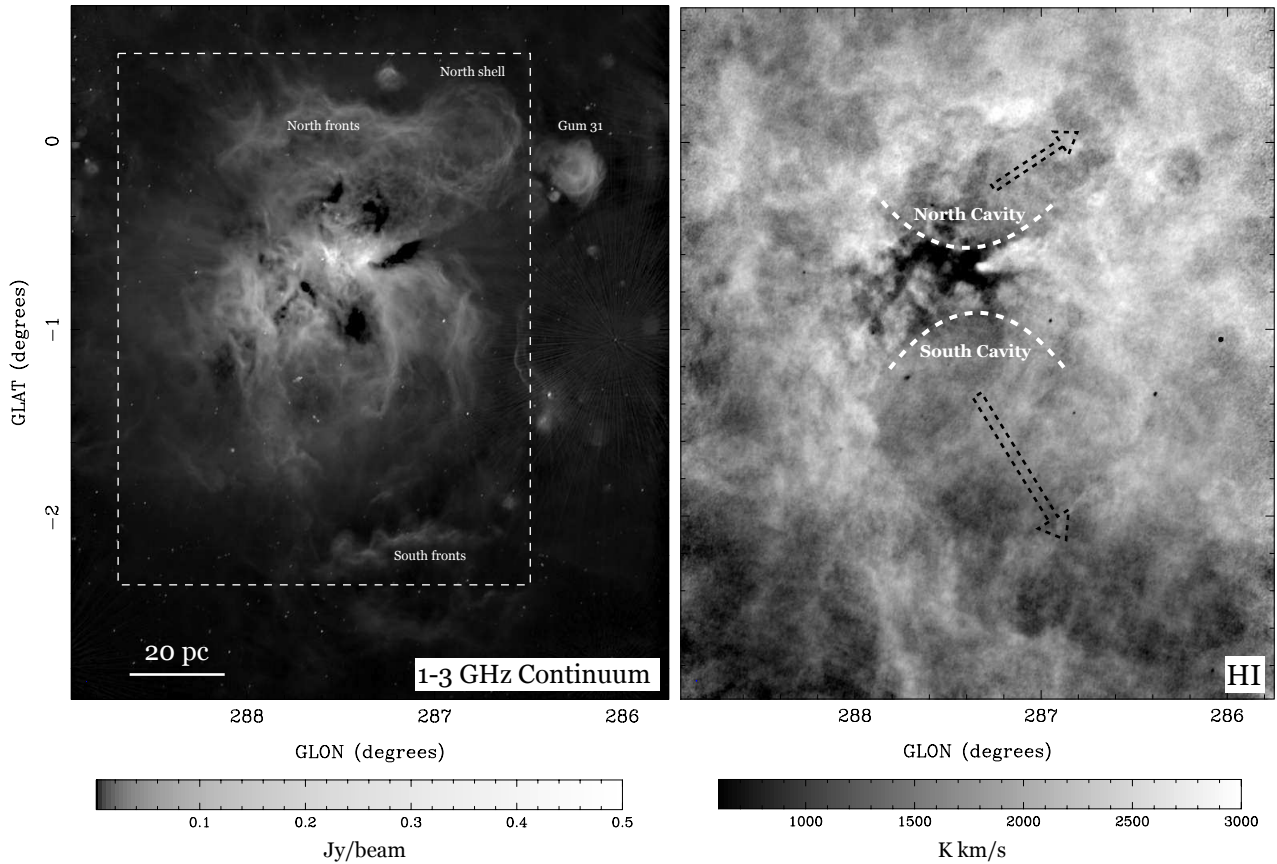,width=0.75\linewidth,angle=90}
\caption{Two products of our ATCA Large Program observing campaign in the CNC-Gum 31 region, the $1-3$ GHz radio continuum image (left) and the \HI\ integrated intensity published in Paper II (right).  The \HI\ map was generated by integrating from $-45\ \kms$ to $0\ \kms$, the velocity range associated with the Carina region (Paper I).  The white dashed-lined rectangle in the left panel shows the region used to estimate the total radio flux and hence derived the ionizing photon luminosity of $Q_\mathrm{H}$.  The curved white dashed lines in the  \HI\ integrated intensity map delineate the south and north cavities discussed in the text.  The bipolar outflow generated by the massive stellar clusters located at the center of the Nebula is clearly detected in the \HI\ map, and they are shown by the open black arrows.  These two images can be compared to further investigate in detail the impact of the stellar feedback in the gas across the Nebula.}
\label{multiphase_hi_cont}
\end{figure*}

Collective feedback from the massive stars have carved two cavities, one in the south and another in the north of the Car \ion{}{1} and Car \ion{}{2} regions.  Although the radio continuum map is not the best tracer to delineate these cavities, we are still able to identify the southern one in our image.  However, the \HI\ map displayed in Figure \ref{multiphase_hi_cont} provides a better view.  These cavities were also identified in the infrared and optical images in \citet{2000ApJ...532L.145S}, and further discussed in \citet{2007MNRAS.379.1279S}.  By examining the morphology of the dust pillars and head-tail structures, they determined the direction of the main source of ionization and stellar winds that created them.  They established that the structures in the cavity in the south point mainly to $\eta$ Carina and Trumpler 16, while the north cavity is mainly shaped by Trumpler 14.  

\subsubsection{North region of the CNC}\label{north-cnc}
In the north of the CNC the radio emission reveals the presence of fronts that point to the center where massive stars are located (Figure \ref{carina_conti_full}).  These fronts are filamentary structures with a morphology roughly perpendicular to the direction of the radiation emanating from the clusters, with a projected distance of $\sim$ 50 pc from west to east.  The fronts seem to be located at different distances from the center, probably due to projection effects.

West from these fronts and centered at ($l$,$b$) $\sim (286.8,0.05)$, we identify a large shell-like structure of radius $\sim$ 10 pc.  There are no stars detected inside the shell that could explain the bubble structure.  According to optical images, tracers such as [\sii] are bright in this region indicating shock excitation (\citealt{2007MNRAS.379.1279S}). \citet{2007MNRAS.379.1279S} suggested the formation of a blister as the most probable scenario that can explain this structure.  In the past, as the fronts were advancing to the north, small cavities were filled with hot plasma from the main \hii\ region.  As the pressure inside the cavity increases, this eventually increases its size, generating a bubble structure.  Comparisons between radio continuum emission and other tracers such as \HI, 8 $\mu$m, and 70 $\mu$m provide further supporting evidence for this scenario, and they will be revisited in Section \ref{multi-phase}.

\subsubsection{South region of the CNC}\label{south_reg_cnc}
We also identify fronts in the southern part of the CNC in the continuum image shown in Figure \ref{carina_conti_full}.  These fronts are weaker in radio emission than the northern ones, and they are located at a larger distance ($\sim$ 60 pc to 80 pc).  These fronts are the final stage of the disruptive feedback from the massive stars that ejected a bipolar outflow creating a large cavity in the gas in the south of the Galactic Plane (Figure \ref{multiphase_hi_cont}).  In fact, as suggested by \citet{2007MNRAS.379.1279S}, these fronts are the southern walls of a bigger bubble of $\sim$ 0.8\degrees diameter.  

\subsubsection{Gum 31}
Farther to the west, the \hii\ region Gum 31 is clearly detected.  This \hii\ region has been studied in X rays (\citealt{2014A&A...564A.120P}), gas tracers (\citealt{2015A&A...582A...2D}), and infrared (\citealt{2013A&A...552A..14O}).   Gum 31 has the stellar cluster NGC 3324 inside, and its shape is a bubble with a radius of $\sim$ 6 pc.  The radio continuum emission has a peak of $\sim$ 0.05 Jy/beam, and it shows multiple layers inside the bubble.  Gum 31 is smaller than Carina, and the gas surrounding the \hii\ region is less disturbed and colder than the gas in the CNC  (Paper I).  The cavity carved by NGC 3324 appears to have a gap in the eastern region, and there is an apparent leakage of ionized gas eastwards, towards the larger northern shell described in Section \ref{north-cnc}.  Two escenarios can explain the bubble structure seen in Gum 31.  One possibility is that the bubble structure is shaped by an expanding shell powered by the stellar cluster NGC 3324.  Alternatively, cloud-cloud collision (CCC) has been suggested as a possible mechanism for the formation of bubble structures in regions of massive star formation (\citealt{1992PASJ...44..203H}; \citealt{2011ApJ...738...46T}; \citealt{2015ApJ...806....7T}).  In fact, \citet{2020PASJ..tmp..238F} has recently proposed that the star formation in the Northern Cloud in the center of the CNC was the result of a CCC.  In this model, a large cloud is impacted by a smaller cloud, creating a compressed layer in the interface between the two structures.  In the highly turbulent compressed layer, massive clumps are formed, allowing subsequent massive star formation to occur.  This mechanism was proposed by \citet{2015ApJ...806....7T} to explain the bubble structure in the RCW 120 region, which is similar in shape to Gum 31.  A detailed study of the gas kinematics will provide a better picture of the mechanism behind the bubble shape identified in images of several gas tracers in Gum 31.

\begin{figure}[tbph]
\centering
\begin{tabular}{cc}
\epsfig{file=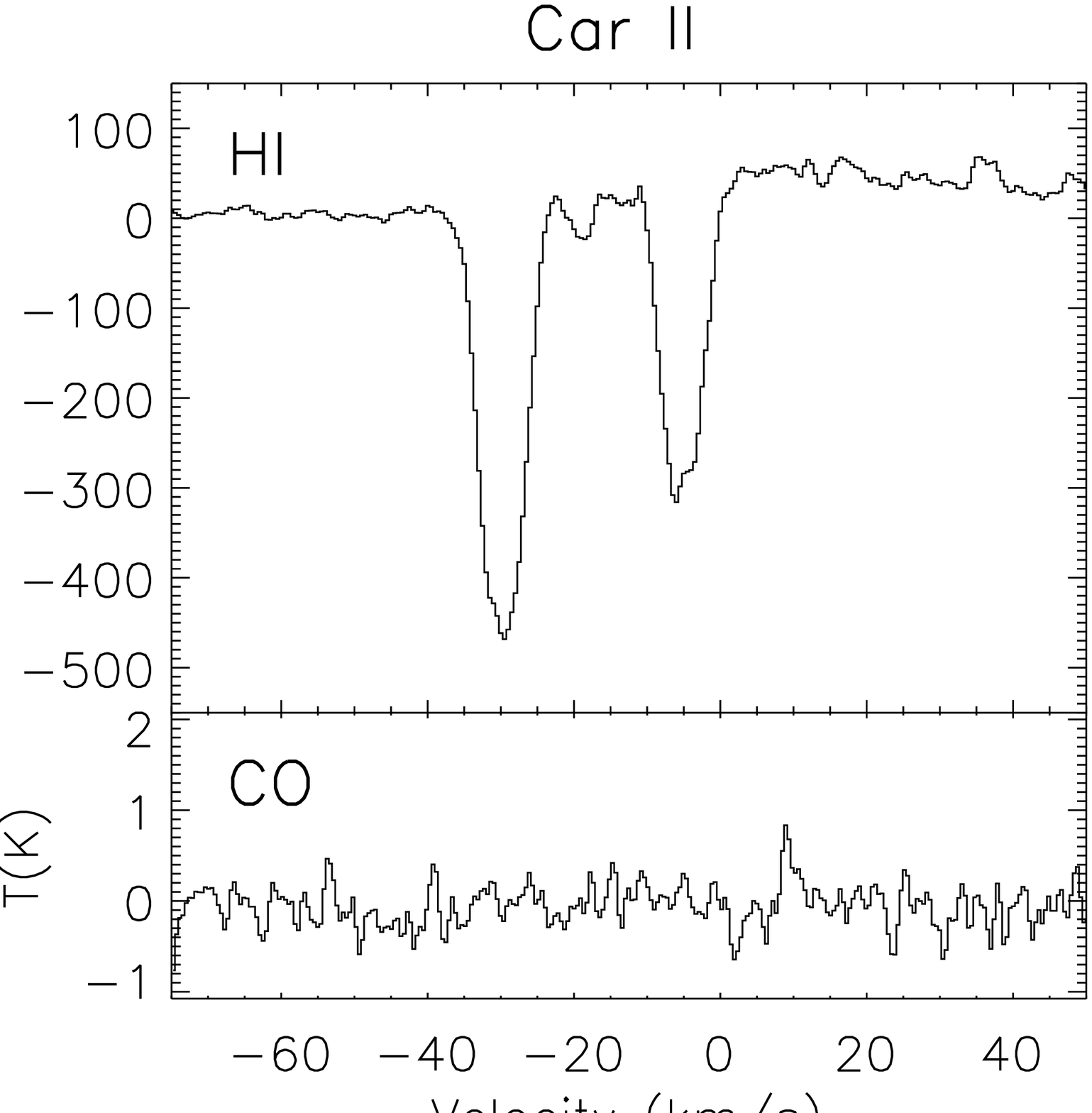,width=0.48\linewidth,angle=0} & \epsfig{file=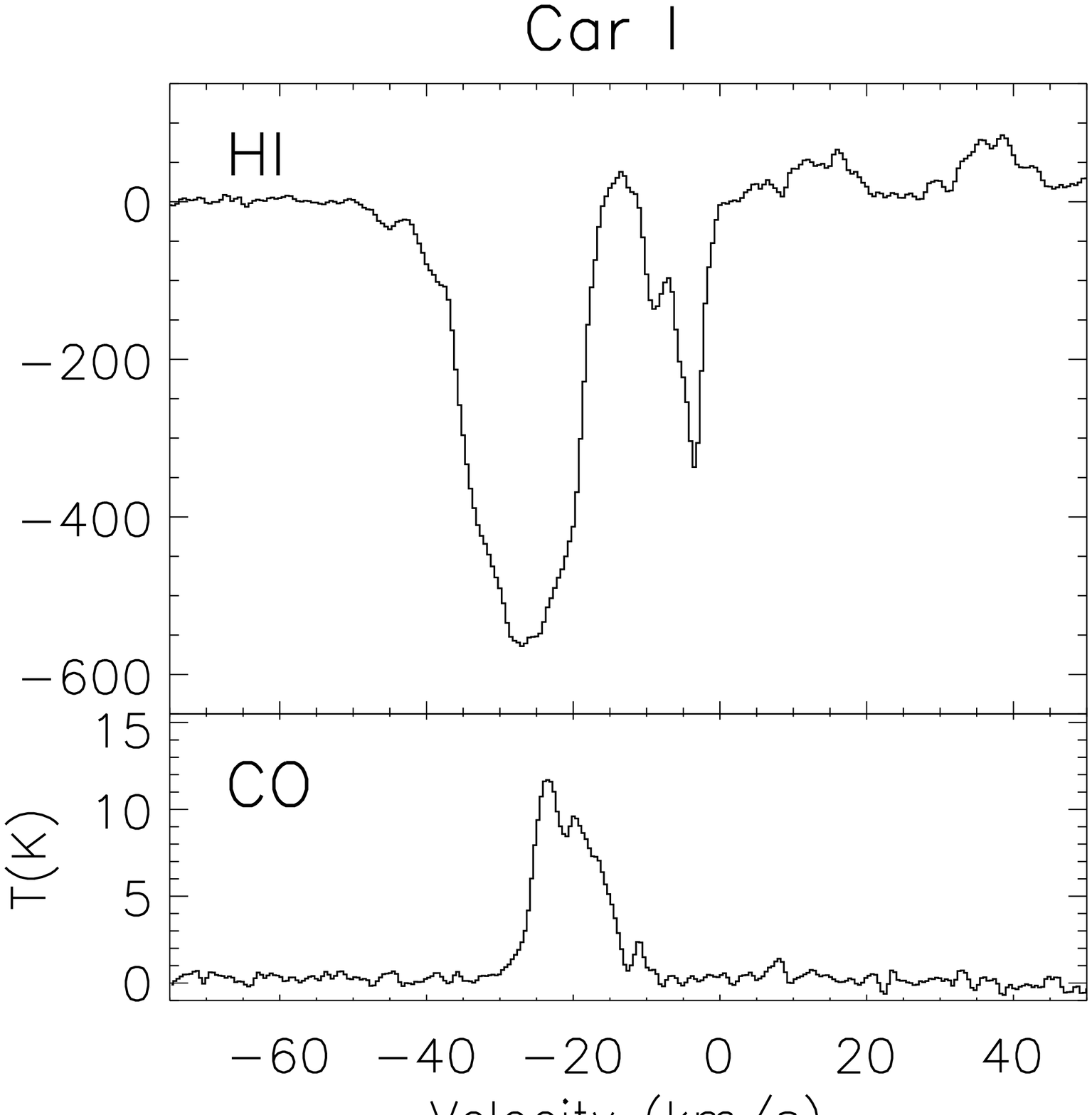,width=0.48\linewidth,angle=0} \\
\epsfig{file=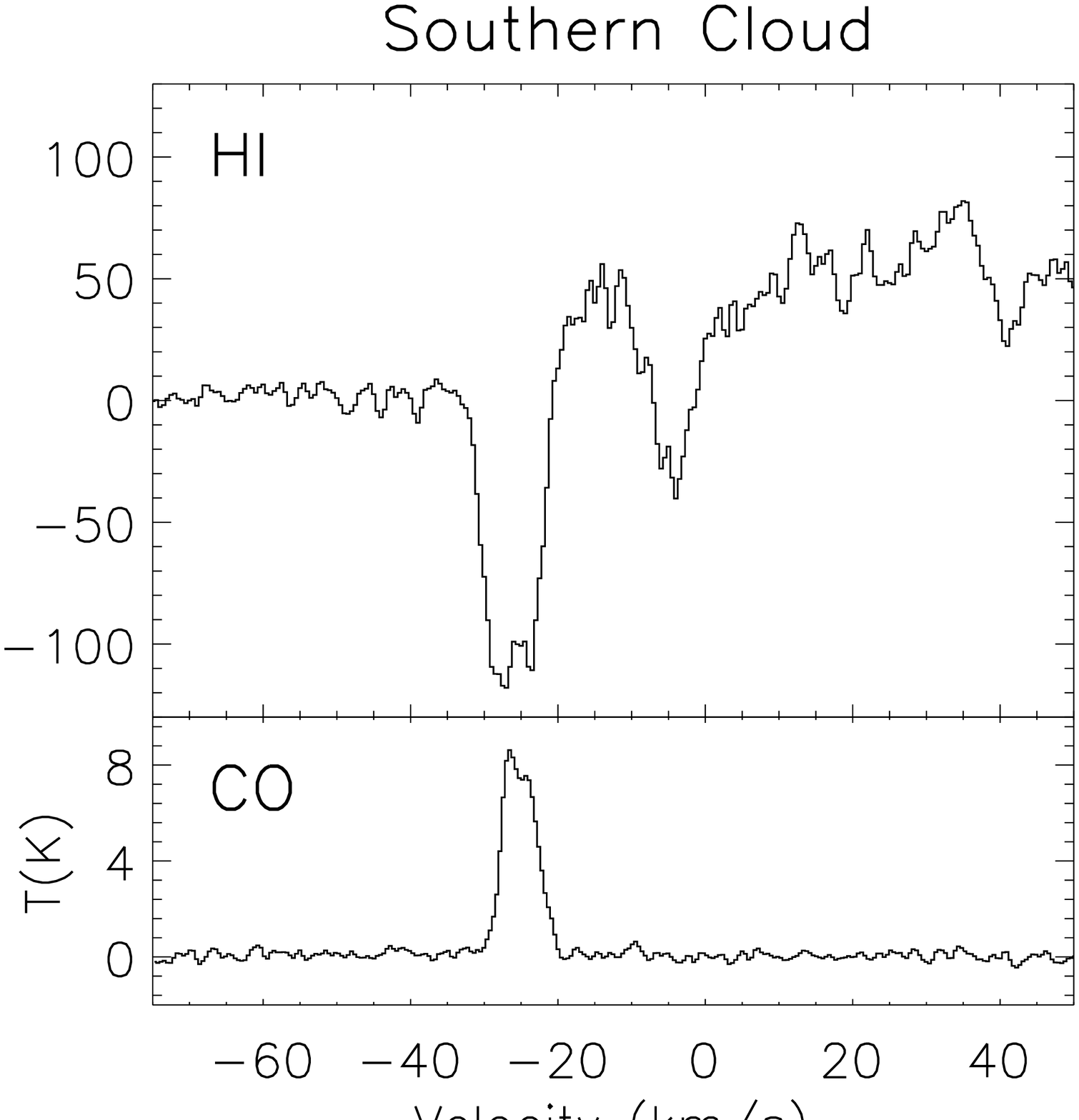,width=0.48\linewidth,angle=0} & \epsfig{file=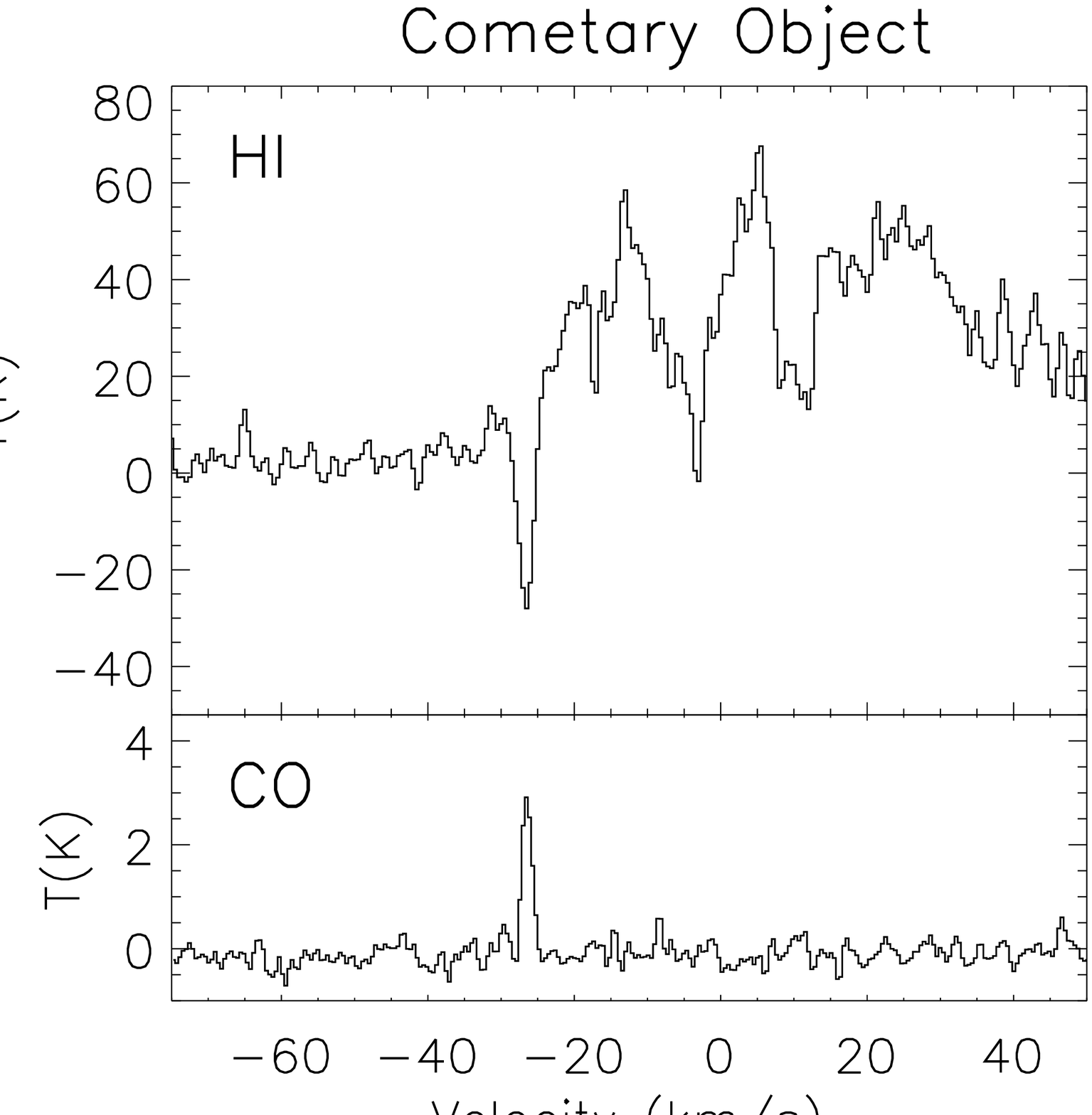,width=0.48\linewidth,angle=0} 
\end{tabular}
\caption{Average spectra from a sample of different regions in the CNC.  In each panel, the upper spectra are from \HI\ 21-cm map, while the lower profiles are from the $\co$ map.  Due to the strong continuum emission, the \HI\ line is seen in absorption in multiple regions.  These absorption features are related to cold/warm atomic gas clouds present in the region.  Some of the velocity components have molecular counterparts, likely to be connected to the denser parts of the cloud that was disrupted by the massive stars.}
\label{spect_hi_cont}
\end{figure}

\section{DISCUSSION}\label{discuss}

\subsection{Muti-phase view of the ISM in the CNC-Gum 31 complex}\label{multi-phase}

\subsubsection{Hydrogen atomic gas}\label{atomic}
Figure \ref{multiphase_hi_cont} shows the two panoramic images produced by our large program with the ATCA.  On the right, we show the integrated intensity map of the the \HI\ 21-cm line (over the velocity range associated with the CNC-Gum 31 region) published in Paper II, and on the left, the radio continuum image presented in this work.  The \HI\ map offers a clearer view of the bipolar outflow discussed in section \ref{morphology}. Towards the north, we see a clear cavity that connects to the North Shell, supporting the scenario that a hot plasma blister is the origin of this bubble structure.  To the south, we see that the fronts from the massive clusters have advanced deep into the southern part of the Galactic Plane, reaching distances larger than 1.5\degrees (60 pc), creating the large cavity seen in the \HI\ map.  This large bubble might be an older version of the shell structure identified in the north of the nebula.  The expansion in the south has been occurring for a longer period of time and is now it has formed a larger structure, and reached larger distances than the northern cavities.

Because of the high brightness of the diffuse radio continuum emission in the CNC, it is extremely difficult to obtain the velocity components of the atomic gas from the line emission profiles. Diffuse continuum emission that is not strong enough to produce \HI\ absorption features can diminish the \HI\ line intensity (\citealt{2015A&A...580A.112B}). However, the \HI\ 21-cm absorption features can be used to obtain the velocity distribution of the cold component of the atomic gas along the light of sight. The \HI\ line will be detected in absorption if the background continuum source brightness temperature is higher than the spin temperature ($T_\mathrm{S}$) of the foreground \HI\ cloud.  Figure \ref{spect_hi_cont} shows a sample of absorption profiles towards four regions in the CNC.  The selected regions includes Car \ion{}{1} and Car \ion{}{2}, the molecular cloud located to the east from Car \ion{}{2} (named Southern Cloud in Paper I), and one of the cometary object presented in Section \ref{particular}.  

\begin{figure}[tbph]
\centering
\epsfig{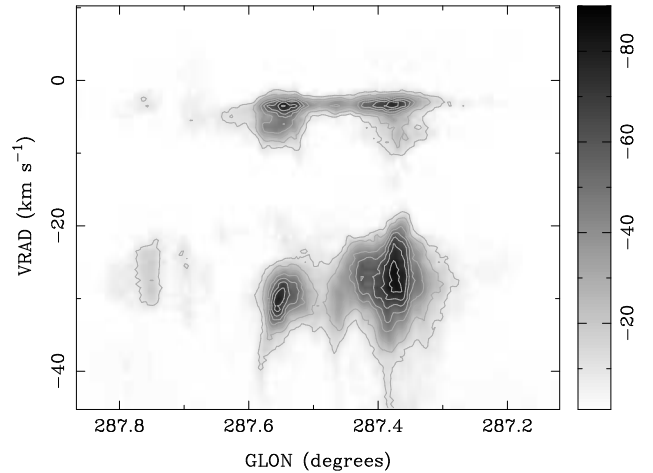}
\caption{Position-velocity (p-v) diagram of the cold component of the hydrogen atomic gas as seen in \HI\ line absorption towards the Car \ion{}{1} and Car \ion{}{2} region.  Contours are $-100,- 90, -80, -70, -60,- 50, -40, -30, -20, -10$ and 0 K. The p-v diagram has been obtained by averaging the \HI\ data cube in the latitude direction, selecting only the negative region of the spectra to extract the absorption components.  The contour bar is in K.  Two cold components are clearly detected along the line of sight in both regions Car \ion{}{1} and Car \ion{}{2}. }
\label{pv_diagram}
\end{figure}

Two main velocity components are clearly identified in the spectra of all the sources, with the strongest at about $-30\ \kms$, and a second component located at about $-5\ \kms$ with a variation of a few $\kms$ between regions.  Previous observations of the H110$\alpha$ recombination line towards Car \ion{}{1} and Car \ion{}{2} revealed a two peak spectrum towards the last one (\citealt{2001MNRAS.327...46B}).  The central velocity of these components were $\sim -35\ \kms$ and $\sim -5\ \kms$, which are coincident with the velocities observed in the \HI\ absorption lines. One possible interpretation of the observed velocities in these tracers, is that the recombination line is tracing two \hii\ regions along the line of sight in the center of the CNC, both being probably powered by the Trumpler 16 and 14 clusters. The \HI\ absorption traces the colder neutral gas that is being pushed by the ionization fronts.  To explore more this idea, in Figure \ref{pv_diagram} we show a position-velocity (p-v) diagram of the cold component of the \HI\ line seen in absorption towards the Car \ion{}{1} and Car \ion{}{2} region.  This figure clearly shows the two absorption components seen in Figure \ref{spect_hi_cont}, and their distribution along the velocity axis.   Each component can be linked to the two expanding shells powered by the Trumpler 16 and 14 clusters.

The \HI\ absorption lines in Car \ion{}{2} do not have CO counterparts.  The massive stars have cleared out the molecular gas in the line of sight towards this region, so molecular line emission is not detected.  On the other hand, CO is clearly detected in Car \ion{}{1}, the Southern Cloud and in the cometary structure at $\sim -35\ \kms$, similar to the velocity of strongest \HI\ absorption feature in the region.  In the case of Car \ion{}{1}, this \hii\ region is located in the line of sight of the Northern Cloud, one of the brightest regions in CO emission in the complex (Paper I).  Additionally, the line width of the absorption line is broader in Car \ion{}{1}.  As has been discussed in previous work, shocks are present in the gas in Car \ion{}{1} (\citealt{2001MNRAS.327...46B}).  High spatial resolution observations of dense molecular tracers with ALMA towards Car \ion{}{1} revealed ionization/shocks fronts that take the shape of arc-like structures (\citealt{2020ApJ...891..113R}).  These structures point to Trumpler 14 as the main source of ionization.  This region will be revisited in Section \ref{NC}.

\begin{figure*}
\centering
\epsfig{file=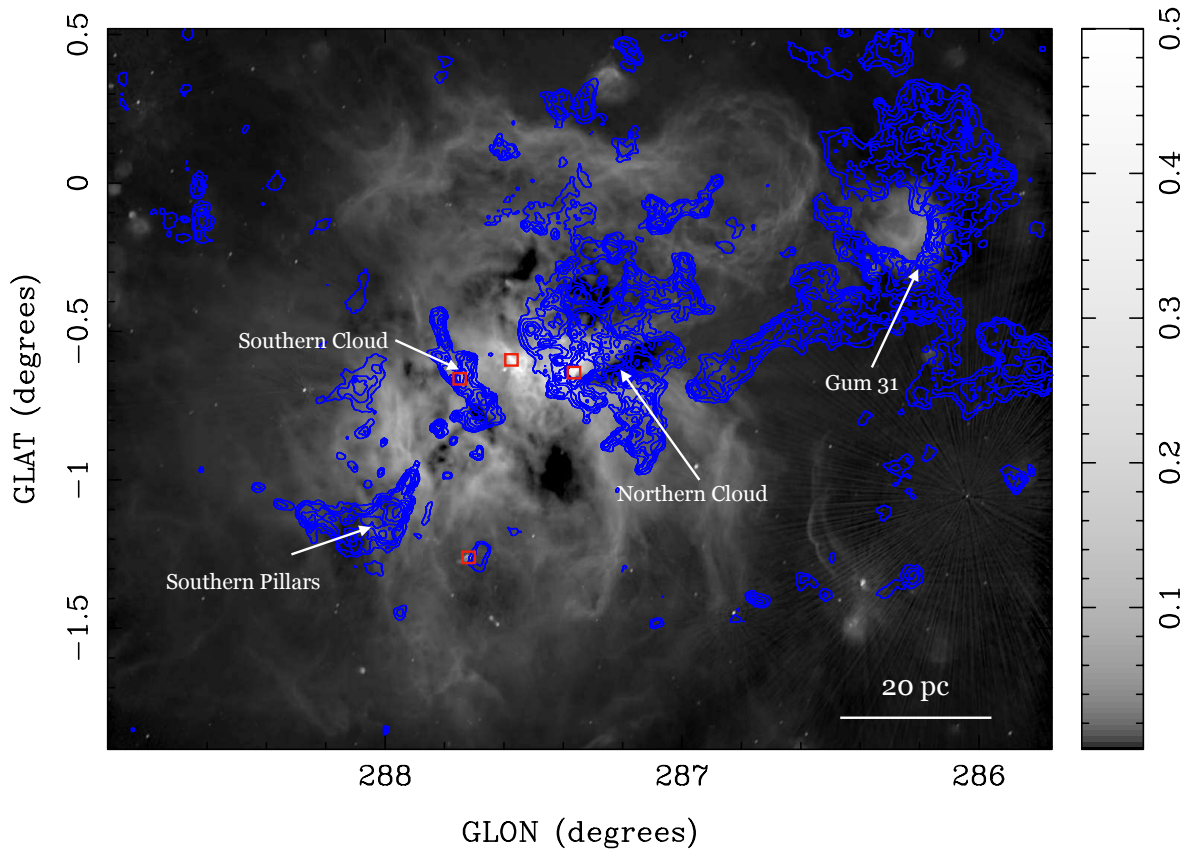,width=0.7\linewidth,angle=90}
\caption{Map of the $\co$ molecular line from Paper I (contour map) overlaid on the $1-3$ GHz radio continuum emission map (background black and white map).  Contour levels are 4, 8 ,16, 25, 36, 49, and 64 K $\kms$.  Color bar is in Jy/beam.  This figure also shows the four main clouds identified in Paper I, the Southern Pillars, Southern Cloud, Northern Cloud and Gum 31.  Red squares show the regions used to extract the spectra shown in Figure \ref{spect_hi_cont}.}
\label{cont_co}
\end{figure*}

Absorption features allow the detection of the colder component of the atomic hydrogen gas (\citealt{2000ApJ...536..756D}; \citealt{2003ApJS..145..329H}; \citealt{2003ApJ...586.1067H}; Paper II), and in some particular cases, the derivation of physical properties such as the optical depth ($\tau$) and $T_\mathrm{S}$ of the \HI\ 21-cm line, two quantities extremely difficult to estimate, but yet of key importance to estimate true atomic gas column densities.  One of the most popular methods to derive these quantities is the ``on-off'' method (\citealt{2000ApJ...536..756D}; \citealt{2003ApJS..145..329H}; \citealt{2003ApJ...586.1067H}; \citealt{2004ApJ...603..560S}; \citealt{2007AJ....134.2252S}).  The technique uses the absorption profile along the line-of-sight from a compact extragalactic continuum source, and a nearby position to estimate the emission profile from the same gas.  By combining both absorption and emission profiles, it is possible to estimate $\tau$ and $T_\mathrm{S}$.  In Paper II, both $\tau$ and $T_\mathrm{s}$ profiles were derived towards the radio source PMN J1032--5917 (\citealt{2007ApJ...663..258B}).  In the Carina-Gum31 region, $\tau \sim 2.1$ and $T_\mathrm{S} \sim$ 100 K for a cloud located at the velocity $-15\ \kms$.  In addition, $\tau$ reaches values $> 3.5$ at the velocity component $0\ \kms$, providing further evidence for the atomic gas being optically thick in regions in the Galactic Plane.  In a future paper, we will implement a detailed modeling of the absorption and emission lines which will allow us to properly correct the effect of the radio continuum emission in the estimate of the \HI\ column density in the CNC.

\subsubsection{Molecular gas}\label{molecular}
Figure \ref{cont_co} shows the distribution of the CO in Carina from Paper I along with the radio continuum emission for comparison.  The spatial distribution map permitted the splitting of the CNC-Gum 31 region into smaller clouds, namely, the Southern Pillars, the Southern Cloud, the Northern Cloud and the Gum 31 region.  Among these regions, the Northern and Southern Clouds are located in regions of strong radio continuum emission in the center of the nebula, which is energy dominated by the Trumpler 14 and 16 clusters.  Although the Southern Pillars are located further away from the center, they are still subject to significant ionization from the clusters.  Indeed, according to the analysis of {\it Spitzer Space Telescope} data made by \citet{2010MNRAS.406..952S}, the pillars structures generally point to the direction of Trumpler 16 and $\eta$ Carina. Stellar feedback has triggered subsequent episodes of star formation propagating across the Southern Pillars, with more than 800 young stellar objects (YSOs) identified in this region, many of them still embedded inside the head of the pillars (\citealt{2010MNRAS.406..952S}).

In Paper I, regional variation of the fraction of molecular gas derived from the CO emission with respect to the total gas mass estimated from the dust emission was clearly detected, with the Southern Pillars and Cloud having smaller percentage of CO ($\sim$ 50\% and 70\% respectively) in their gas compared to the Northern Cloud and Gum 31 that have more than 80\%.  The most interesting case among all the clouds in the region is the Northern Cloud.  This cloud is located close to the star cluster Trumpler 14 and nearby Trumpler 16, but the molecular mass fraction is still comparable to the fraction observed in Gum 31, a region quite isolated from the strong radiation fields of the massive star clusters in the CNC. This might be the result of initial density inhomogeneities in the large parent cloud that gave birth to the first generation of stars in Carina.  If the gas density was lower towards the Southern Pillars, then the ionization/shock fronts of newly formed stars travelled further in that direction, shaping the pillars.  On the other hand, regions of higher density can shield more efficiently against ionization fronts, allowing regions such as the Northern Cloud to survive and maintain high levels of molecular gas fraction.  To support this explanation, there is a noticeable difference in the number of YSOs detected in the Southern Pillars (> 800) and in the western region where the Northern Cloud is ($\sim$ 100), suggesting a lower level of triggered star formation towards the direction of the denser gas (\citealt{2010MNRAS.406..952S}).

\begin{figure*}
\centering
\epsfig{file=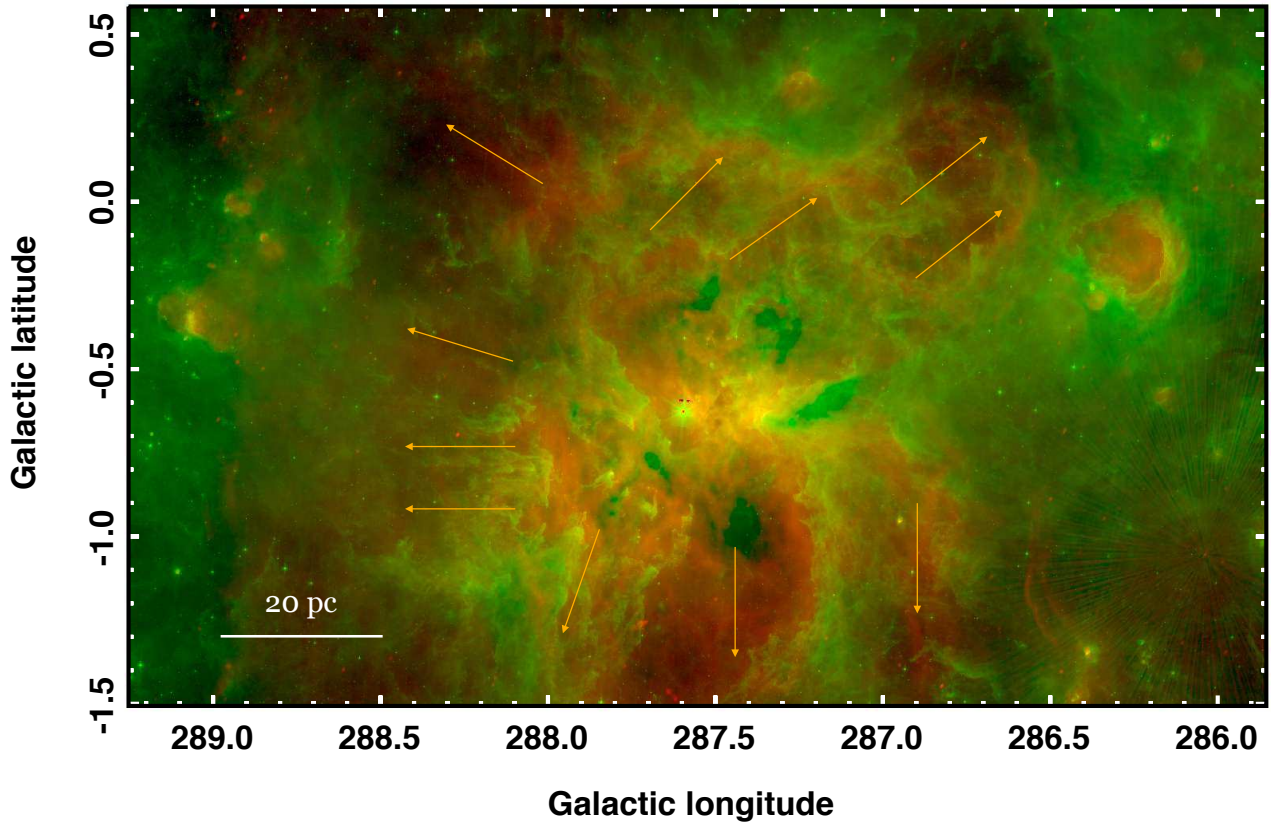,width=0.75\linewidth,angle=90} 
\caption{Color composite image of the CNC-Gum 31 complex.  Green shows the {\it Spitzer} IRAC 8 $\mu$m map, while red shows the ATCA continuum image.  A clear spatial anti-correlation between the two tracers is observed, with the radio continuum emission filling the cavities observed in the 8 $\mu$m map. The 8 $\mu$m map traces the PAH emission, so it will be bright on the cloud's surfaces. Orange arrows show the apparent direction of the hot gas traced by the radio continuum map as it permeates through the denser gas, illuminating the cloud surfaces as traced by PAH emission.}
\label{cont_8um}
\end{figure*}

\subsubsection{8 $\mu$m}\label{8um}
Figure \ref{cont_8um} shows a comparison between the radio continuum emission and the 8 $\mu$m image from Spitzer.  In general, we detect some anti-correlation between the two tracers, with the radio continuum emission filling the cavities observed in the 8 $\mu$m map.  A Spearman's rank correlation between the two images gives a value of 0.42, which indicates a poor level of correlation.  As 8 $\mu$m traces the polycyclic aromatic hydrocarbon (PAH) emission, it will be bright on the surfaces of the clouds being radiated by the nearby massive stars. Now, with the radio continuum map in hand, we are able to witness the complex distribution of the different gas phases in Carina, which deviates strongly from the classical view of a spherical region expanding symmetrically into the surrounding ISM.  Because of the to exquisite detail of our radio continuum image, we can see that the hot gas can indeed permeate through the molecular cloud, and shape the material into features such as pillars, small shells and arc-like structures.  

As the ionization fronts advance thorough the cloud, they can penetrate deep towards low density regions, and ultimately escape.  Figure \ref{carina_conti_full} showed structures resembling ``plumes'' at the edges of the \hii\ region.  These features are likely to be tracing hot gas that had eroded the material in the nebula, extending to outside the molecular cloud, and now diffusing with the global component of the hot gas in the Galactic Plane at distances > 100 pc.  There is clear alignment between the morphology of the illuminated gas traced by PAH emission, and the plume-shaped structures traced by the radio continuum map.  The direction of the elongated structures identified in the north-east section of the 8 $\mu$m map (such as pillars and walls) is roughly the same as the direction of the plumes detected in the radio continuum (Figure \ref{carina_conti_full}). This correlation is more evident towards the Southern Pillars region, where the radio continuum emission is distributed along the cavities between the pillars.  

\begin{figure*}
\centering
\epsfig{file=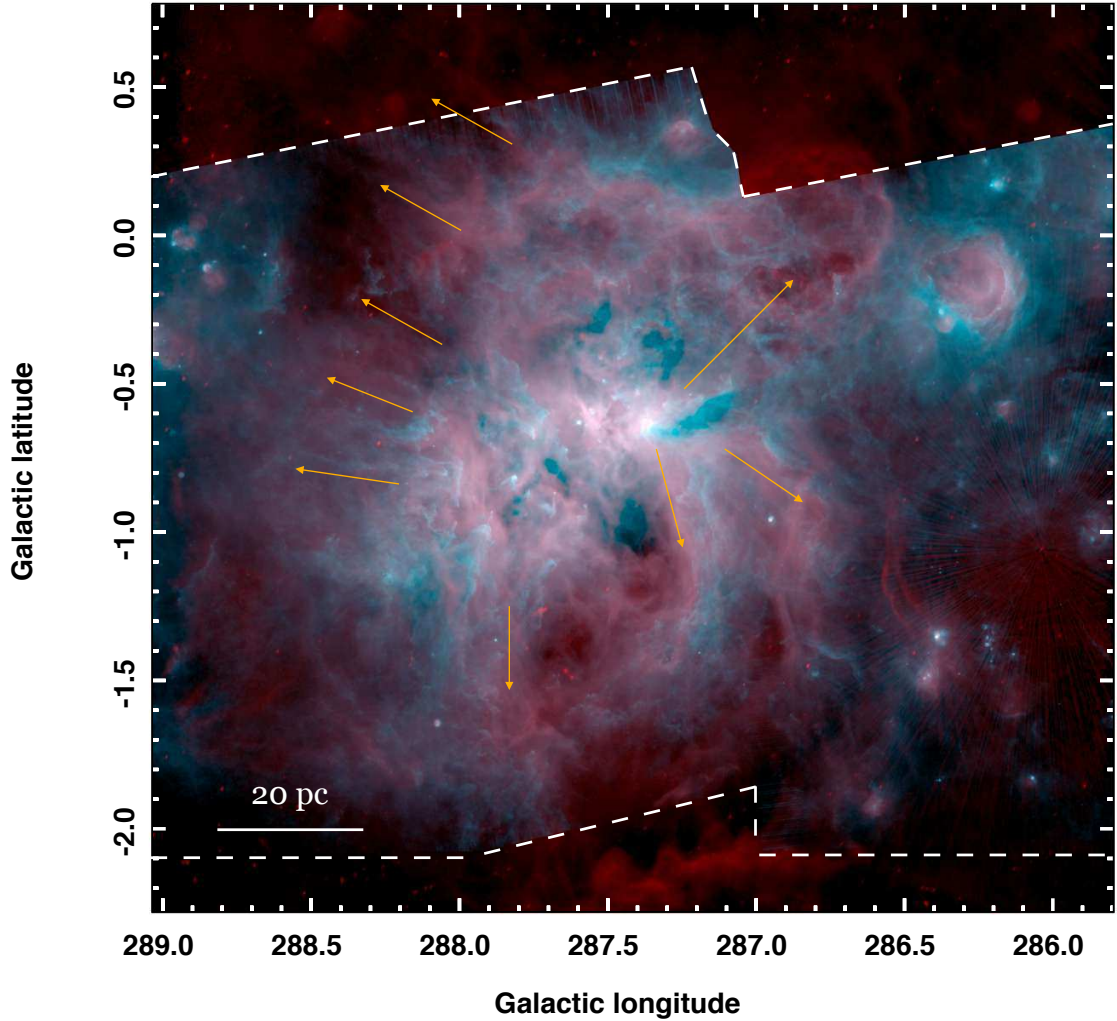,width=0.8\linewidth,angle=90}
\caption{Color composite image of the CNC-Gum 31 complex.  Cyan shows the {\it Herschel} 70 $\mu$m map, while red shows the ATCA continuum image.  The dashed white line shows the 70 $\mu$m map coverage.  Pink color shows regions where both 70 $\mu$m and radio continuum map emission are present, marking the location where dust has been locally warmed by the radiation field.  The orange arrows show the approximate direction of the hot gas as it permeates through the CNC molecular complex, and escapes towards the Galactic Plane.}
\label{cont_70um}
\end{figure*}

\subsubsection{70 $\mu$m}\label{70um}
In Figure \ref{cont_70um} we show a color composite image of the {\it Herschel} 70 $\mu$m and our radio continuum image.  As distinct from the comparison discussed in Section \ref{8um}, some spatial correlation between the 70 $\mu$m and radio continuum map is clearly identified.  A Spearman's rank correlation between these two images gives a value of 0.75, which indicates a higher level of correlation than the estimated between the radio continuum  and the 8 $\mu$m maps.  A possible interpretation of this result is related to the {\it in situ} heating of dust particles in \hii\ regions (\citealt{2007MNRAS.379.1279S}).  The dust in the material inside \hii\ regions can be efficiently heated by the emission from the ionized gas.  Thus, the warm gas traced by 70 $\mu$m will be spatially coincident with regions of strong continuum emission as is shown on Figure \ref{cont_70um}. Using coarser spatial resolution images, \citet{2007MNRAS.379.1279S} fitted a global spectral energy distribution (SED) to the CNC using wavelengths from the infrared to the radio.  They successfully recovered the SED profile using three optically thin grey body components with temperatures of 30, 80 and 220 K (although the last component was likely to be related to emission from PAH and silicates rather than dust continuum emission).  By contrast, in Paper I we derived a dust temperature map of the CNC-Gum 31 complex by making a pixel by pixel SED fitting to the Herschel maps, obtaining values $\sim 30-40$ K in the central region of the nebula.  In our analysis, we only used a single grey body component which was sensitive mainly to the cold dust component, similar to the colder component in \citet{2007MNRAS.379.1279S} analysis. \citet{1978ppim.book.....S} estimated that the temperature equilibrium between absorption of trapped Ly$\alpha$ photons and infrared emission will be achieved at $\sim$ 80 K, assuming solar metallicity and a gas density of $\sim 10^3$ cm$^{-3}$.  Based on this result, \citet{2007MNRAS.379.1279S} suggested that a population of warm dust is intermixed with the ionized gas in several regions in the CNC, particularly in the gas surrounding the surfaces of the clouds.  However, a complete calculation of the grain equilibrium temperature in the CNC region would require a better assessment of the relative contribution to the grain energy gain from different sources such as the $L_\mathrm{\alpha}$ photons, the direct heating by stellar photons, and collision of grain with the gas.  Direct heating by stellar photons can be important in the vicinity of the massive stellar clusters, while the gas collision can become important in region of high density gas, both of which are possible in Carina.  This type of analysis is beyond the scope of this paper, and will revisited in future publications.

\begin{figure*}[tbph]
\centering
\epsfig{file=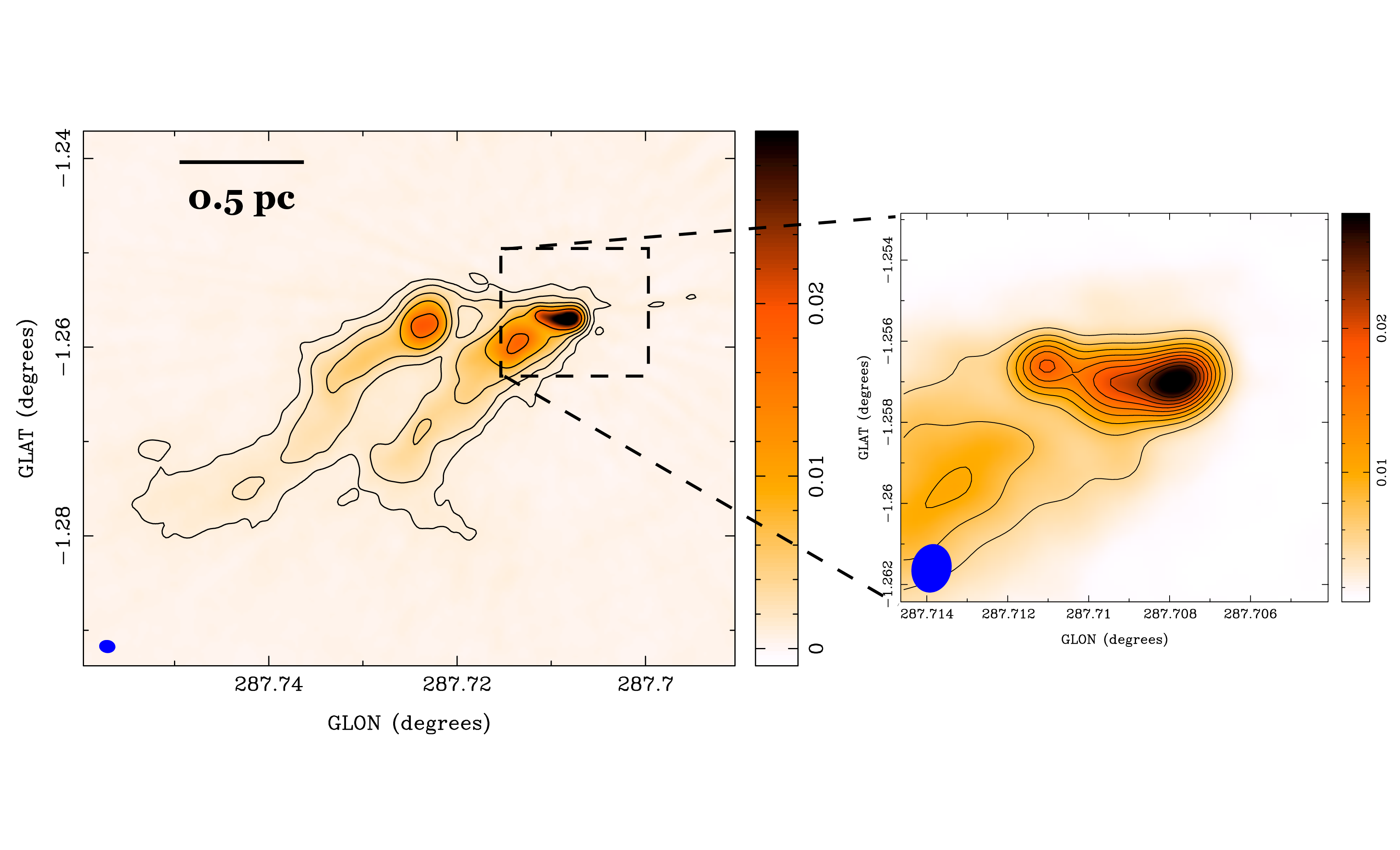,width=0.55\linewidth,angle=-90}
\caption{Left: Closer view of $1-3$ GHz radio continuum map towards the cometary structures identified in the center of CNC.  The color bar is in Jy/beam.  The beam size in the map is $5\farcs8 \times 4\farcs5$.  Two cometary structures are clearly identified, with the brightest emission located at the heads.  Right:  Zoomed view of the brightest head.  The beam size of the image is $4\farcs3 \times 3\farcs5$.  The higher spatial resolution map reveals two structures, probably related to an onsite star formation region.}
\label{cometary_cont}
\end{figure*}

Figure \ref{cont_70um} also shows the plume features identified in Figures \ref{carina_conti_full} and Figure \ref{cont_8um}.  This is expected because the hot gas will heat the dust as it travels across the low density material between the denser gas that composes structures such as pillars (seen in the Southern Pillars region) and walls (seen in the Northern Cloud region).

\subsection{Individual regions}\label{particular}
In this section we will discuss in detail a few special regions identified in the ATCA radio continuum image.  Different from the imaging parameters used in Figure \ref{carina_conti_full}, in this section we use a robust parameter equal to $-0.5$ in order to increase the spatial resolution of the generated maps.  As reported in Section \ref{imaging}, the beam size of the global radio continuum map shown in previous sections is 24\farcs4 $\times$ 15\farcs8.  We have improved the spatial resolution of the radio continuum maps by a factor of $\sim$ 3.5, allowing us to obtain higher level of details present in the maps of each of the individual structures discussed here.

\subsubsection{Cometary Structures}\label{cometary}
Two cometary objects are clearly identified in the southern region of the map shown in Figure \ref{carina_conti_cent}.  The structures seem to point to the massive stars in Collinder 228, although they might also be affected by Trumpler 16.  These objects are not detected in H$\alpha$, or infrared tracers such as 8 $\mu$m and 70 $\mu$m.  Additionally, the \HI\ spectra along the line of sight to these cometary objects shows multiple absorption features across the velocity associated with Carina: $-26\ \kms$ and $-5\ \kms$, and beyond at $17\ \kms$ (Figure \ref{spect_hi_cont}).  We also detect CO emission along the line of sight towards these structures.  The velocity component in the CO shows a clear peak at $\sim -26\ \kms$, similar to one of the absorption HI components.  This suggests the presence of a dense gas cloud along the line of sight towards the cometary structures.  However, given the multiple velocity components detected in the HI absorption spectrum, we suspect that these cometary structures are probably located behind the atomic gas that fill the southern cavity.

\begin{figure*}[tbph]
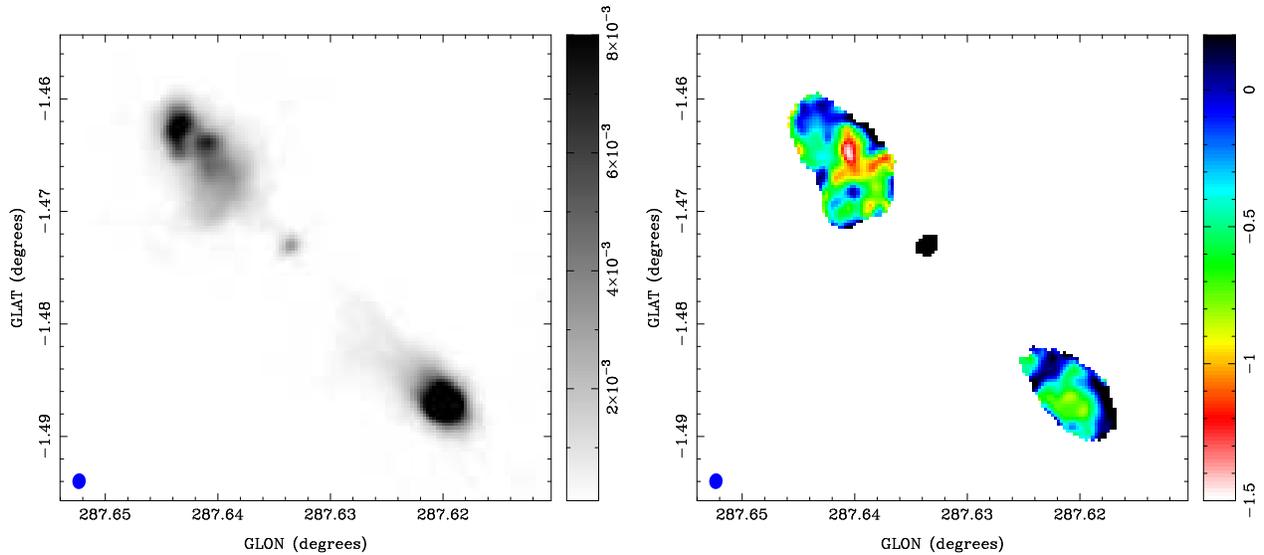

\centering
\begin{tabular}{cc}
\epsfig{file=Figure13_a.ps,width=0.45\linewidth,angle=0}  & 
 \epsfig{file=Figure13_b.ps,width=0.45\linewidth,angle=0}
\end{tabular}
\caption{Left: $1-3$ GHz radio continuum emission map of the object with an identified  bipolar outflow/jet structure.  The image was created using multi-terms and wide frequency imaging with the CASA task {\it tclean}.  The color bar is in Jy/beam.  We identified a clear central source, with a bipolar emission structure, similar to FR II extragalactic sources. Right: Spectral index map of the region.  The values are > $-0.5$ in the lobes, which are consistent with non-thermal emission in those regions.}
\label{jets_image}
\end{figure*}

Figure \ref{cometary_cont} shows a closer view of the $1-3$ GHz radio continuum map towards the cometary structures.  The beam size of the map is $5\farcs8 \times 4\farcs5$, which offered a better view of the internal structure.  We clearly identify the two structures, with their heads and tails.  A higher spatial resolution radio continuum map shows that the brightest structure has two components, probably a bipolar structure related to an onsite star formation event.  We suggest that these regions correspond to the latest stages of triggered star formation happening in some regions in Carina.  It has been established with relatively robustness that the current star formation in Carina has been triggered by previous generation of massive OB stars (\citealt{2002MNRAS.331...85R}; \citealt{2004A&A...418..563R}; \citealt{2010MNRAS.406..952S}).  Different regions in the nebula might be in different phases of evolution depending on how the star formation in that particular region has been induced by the stellar feedback.  According to \citet{2010MNRAS.406..952S}, this process can be separated into three main stages.  In the first stage, the stellar feedback from the previous generation of massive stars have ionized and pushed back the lower density material around massive and dense gas clumps, leaving an ionization shadow behind them.  In this initial stage, some Class 0/I YSOs might be already present, but still embedded inside the clumps.  The time scale for this stage is $\sim$ 10$^5$ yrs. This is probably the case for the main pillars in the Southern Pillars region in Carina.  In the second stage, which happens after a few 10$^5$ yrs, the pillars are even more eroded and accelerated backwards by the feedback.  Due to the continuous externally exerted pressure, they will continue forming a second generation of stars.  The stellar population from stage one is no longer embedded, and in the process of disperse after the gas in the clump has been removed.  In the last stage, once the region is $\sim$ 10$^6$ yrs old, the clump now has a structure more similar to a cometary cloud, with little gas remaining due to the continuous erosion from the nearby stars.  Probably, the cometary clouds we detected in the radio continuum map correspond to the last stage, where almost no gas remains, and isolated structures resembling small pillars or cometary clouds are the only survivors of previous multiple episodes of star formation.

\subsubsection{Bipolar jet/outflow structure}
Slightly below the cometary structures, we identified a source with structure similar to a bipolar jet or outflow.  Figure \ref{jets_image} offers a high spatial resolution image of the $1-3$ GHz radio continuum emission towards this object.  In this case, the imaging was performed using the multi-term and multi-frequency (MT-MFS) algorithm available with the CASA task {\it tclean}.  MT-MFS allows us to maximize the uv coverage by using the full $1-3$ GHz wide spectral coverage available from our ATCA observations. MT-MFS provides the spectral shape of the source by modeling the spectrum of each flux component by a Taylor series expansion about a reference frequency.  In our case, we have used two Taylor terms and a robust parameter equal to $-0.5$ in the imaging process.  The resulting spatial resolution of the map was $5\farcs0 \times 4\farcs3$.  

Figure \ref{jets_image} reveals a central source, with two lobes extending to oposite directions.  The projected size of the object is $\sim 2.3\arcmin$ measured from one extreme to the other.  The lobe extending to the north-east shows two bright spots at the very end of the jet structure, with a maximum brightness of 9.5 mJy/beam.  On the other hand, the lobe extending to the south-west shows a single bright spot, with a peak brightness of 26.7 mJy/beam.  We suspect that this object is an extragalactic radio source not previously identified, as the radio extragalactic surveys usually avoid the Galactic Plane.  This radio source could be a radio galaxy or a quasar.  According to \citet{1974MNRAS.167P..31F}, these objects can be classified into two groups based on the relative distance of the brightest emission on opposite sites of the central source with respect to the lowest emission level.  Fanaroff-Riley Class I (FR I) objects have the brightest spots located closer to the central source than the low level emission regions.  On the other hand, FR II objects have their bright spots located at the extreme ends of the lobes.  According to this classification, the source identified in Figure \ref{jets_image} seems to show an emission morphology similar to a FR II object.  

Figure \ref{jets_image} also shows a spectral index map of the source.  Our data allows us to estimate the spectral index values in the lobes only, as the emission in the jets are relatively weaker.  The range of the values are $> -0.5$ in the lobes, with some regions showing values close to $-1$ in the lobe extending to the north-east.  These are values consistent with non-thermal emission, typical of FR II objects.  Follow up observations to this interesting object will clarify its nature, and will provide further details on its physical properties.

\subsubsection{Detailed view of Car \ion{}{1}}\label{NC}
Trumpler 14 and the bright \hii\ region Car \ion{}{1} have been the target of numerous studies at multiple wavelengths.  The extreme-ultraviolet photons released by Trumpler 14 ionize the ambient neutral media, and play an important role in the heating and chemistry of the gas, making this target ideal to study the physical condition of PDRs and the FUV radiation field (\citealt{2002MNRAS.331...85R}; \citealt{2003A&A...412..751B}; \citealt{2005ApJ...634..436B}; \citealt{2018A&A...618A..53W}).  \citet{2019ApJ...878..120S} observed the 158 $\mu$m line of [\ion{C}{2}] in the gas nearby these regions using the Stratospheric Terahertz Observatory 2 (STO2).  They found that the bright [\ion{C}{2}] emission traces the PDR and the ionization fronts in Car \ion{}{1}.  Comparison of [\ion{C}{2}] with other gas tracers such as \HI\ 21 cm, \co\ and radio recombination lines, revealed that the \hii\ region is expanding towards us, and that the destruction of the molecular cloud is driven by UV photo evaporation. 

\begin{figure*}
\centering
\epsfig{file=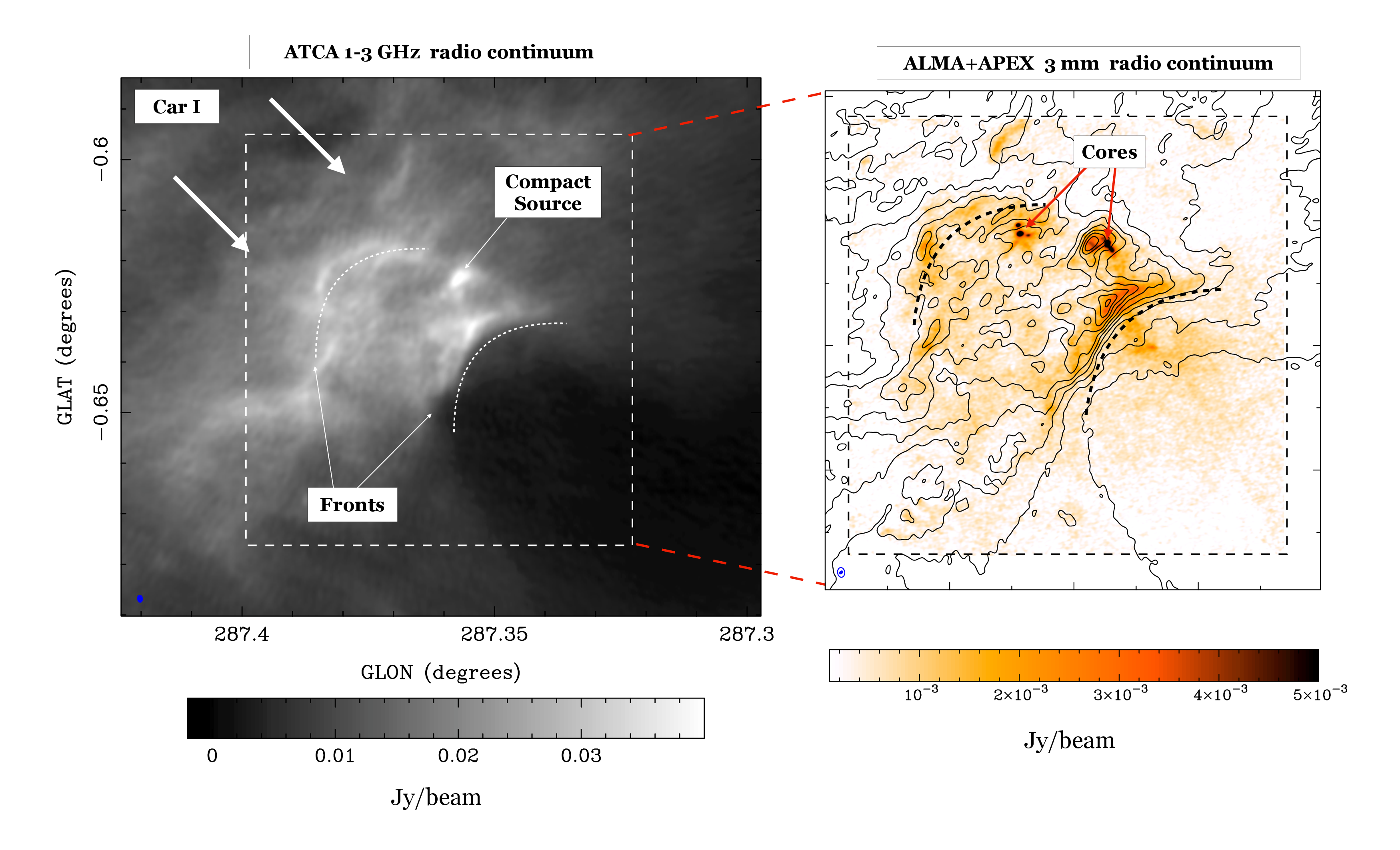,width=0.55\linewidth,angle=90}
\caption{Left: ATCA $1-3$ GHz radio continuum image towards the \hii\ region Car \ion{}{1}.  This image was generated with robust parameter equal to $-0.5$ to maximize the spatial resolution.  The beam size is $5\farcs6 \times 4\farcs1$.  White arrows show the approximate direction of the radiation from massive stars in the Trumpler 14 cluster.  The white dashed arcs show the regions where we identified ionization fronts in ALMA 3 mm continuum data (\citealt{2020ApJ...891..113R}).  Right: ALMA + APEX combined continuum image at 3 mm from \citet{2020ApJ...891..113R}.  Black contours shows the $1-3$ GHz radio continuum emission, with levels of 4 mJy/beam $\times k$ starting at 4 mJy/beam, with $k =$ 1, 2, 3, 4, 5, 6, 7, 8, and 9.  The black dashed box shows the region covered by ALMA observations.  The black dashed arcs shows the ionization front in left panel.  Red arrows shows the position of the resolved cores identified in our ALMA 3 mm continuum data.}
\label{NC_images}
\end{figure*}

In \citet{2020ApJ...891..113R} we presented high resolution maps ($\theta_\mathrm{fwhm} \sim 2\farcs3$) towards two regions with extremely different physical properties in the CNC using ALMA at 3 mm.  We targeted the continuum emission along with a sample of molecular emission lines such as \hcop, \hcn, and \sio, among others.  One of the targeted region was Car \ion{}{1}, which is located in the Northern Cloud.  This region is being severally affected by stellar feedback from the nearby high-mass stars in Trumpler 14.  The other observed region was located farther away from the massive stars, more precisely in the Southern Pillars region, and thus less disturbed by the stellar feedback.  Our study revealed that the region in Car \ion{}{1} has less but more massive cores than the region located in the pillars, providing evidence towards a prominent role of the stellar feedback in the fragmentation process inside clumps.  Now, with the $1-3$ GHz radio continuum image in hand, it is possible to complement this view with the information provided by the ionized gas.

Figure \ref{NC_images} shows a $1-3$ GHz radio continuum map towards the Car \ion{}{1} region.  As in the two previous sections, a robust parameter equal to $-0.5$ has been used to increase the spatial resolution.  The achieved beam size of the map is $5\farcs6 \times 4\farcs1$.  An unprecedented detailed radio continuum map of Car \ion{}{1} is revealed, with multiple features such as filaments and arc-like structures.  \citet{2001MNRAS.327...46B} obtained a 4.8 GHz continuum map of Car \ion{}{1} with similar spatial resolution, but our map is more extended and with better image fidelity due to our superior $uv$-coverage.  The two ionization fronts previously detected at 4.8 GHz by  \citet{2001MNRAS.327...46B}  and by \citet{2020ApJ...891..113R} at 3 mm are detected with even more detail in our $1-3$ radio continuum map.  The peak radio continuum emission at $1-3$ GHz in Car \ion{}{1} is located in a compact source, with a value of 43 mJy/beam.  This radio continuum compact source is located close to one of the massive cores identified in the ALMA 3 mm data.  Figure \ref{NC_images} also shows the ALMA 3 mm continuum map, which has been combined with a frequency extrapolated APEX map to recover zero $uv$-spacing flux (\citealt{2020ApJ...891..113R}).  The diffuse emission shows in general a good spatial correlation between the $1-3$ GHz radio continuum map and the 3 mm continuum image.  The notorious exceptions are the compact sources identified as cores in the 3 mm map in \citet{2020ApJ...891..113R}, and some diffuse emission present at 3 mm, but not at $1-3$ GHz.  However, this diffuse continuum emission at 3 mm correlates well with the molecular gas traced by \hcop, suggesting that this component traces the warm dust emission rather than the ionized gas.  As discussed in Section \ref{70um}, this represents another evidence of {\it in situ} heating of dust particles in \hii\ regions.  In Car \ion{}{1}, the ionized gas and the warmed dust co-exist spatially, and the ALMA 3 mm map traces both components.

\begin{figure*}
\centering
\epsfig{file=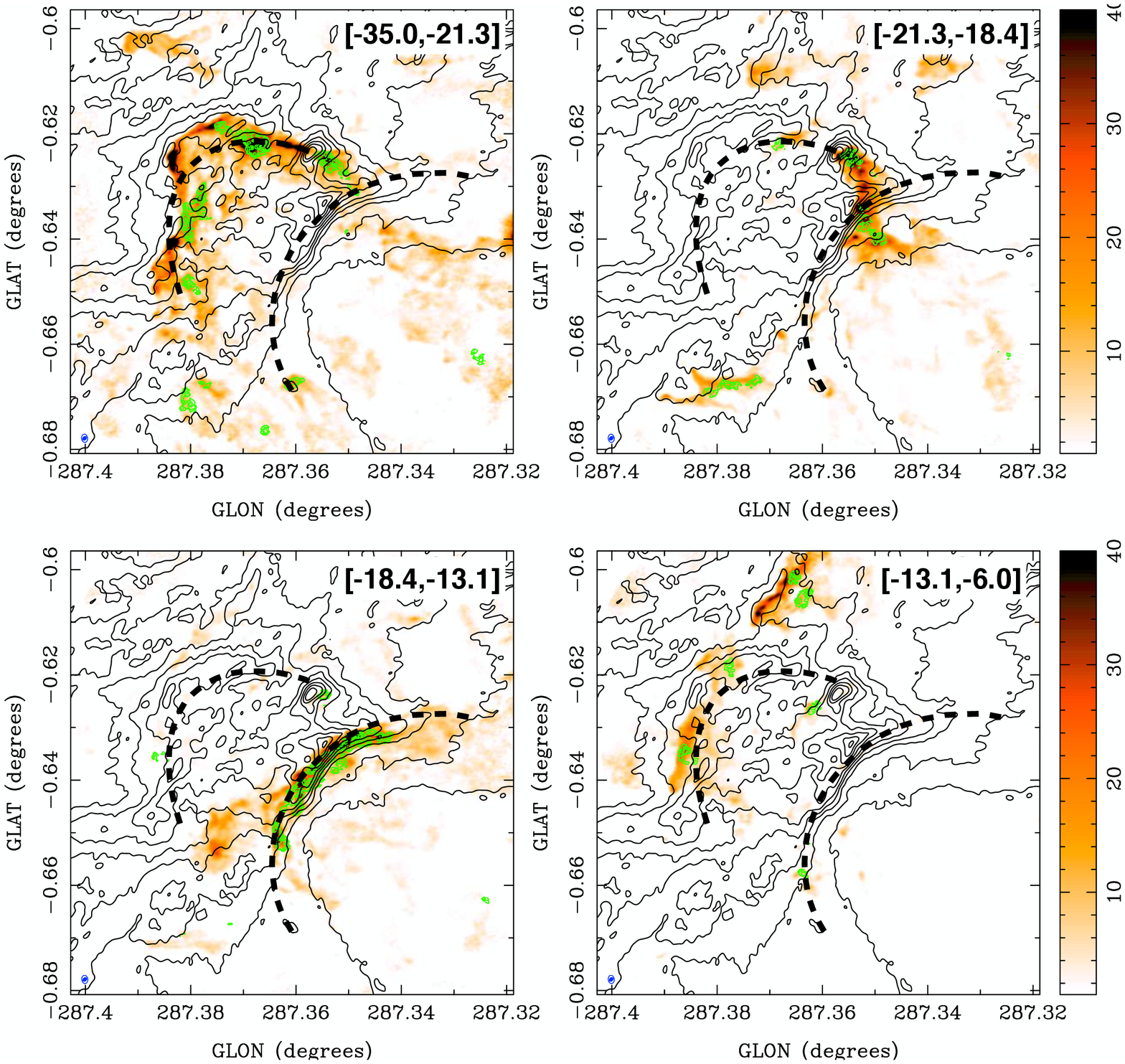,width=0.7\linewidth,angle=90}
\caption{Comparison between the $\hcop$ line obtained with ALMA (color map) and the ATCA $1-3$ GHz radio continuum (black contours) map in the Car \ion{}{1} region. Each panel shows the $\hcop$ intensity map integrated over the velocity range shown in the upper right corner in $\kms$ (\citealt{2020ApJ...891..113R}).  As in Figure \ref{NC_images}, black contours shows the ATCA $1-3$ GHz radio continuum, and black dashed arcs shows the ionization fronts. Green contours show ALMA SiO intensity map integrated over the corresponding velocity ranges, with levels equal to 0.5, 1, 2, 4, 8, and 16 K\ \kms\ (\citealt{2020ApJ...891..113R}).}
\label{NC_images_velo}
\end{figure*}

Figure \ref{NC_images_velo} shows high spatial resolution maps of the \hcop\ and \sio\ lines towards Car \ion{}{1} obtained with ALMA (\citealt{2020ApJ...891..113R}).  The velocity distribution of these tracers have been decomposed into four main components, mainly based on the shape of the different fronts distinguishable at each velocity range.  The distribution of the \hcop\ line in Car \ion{}{1} follows the structure of the fronts traced by the $1-3$ GHz radio continuum image.  \citet{2020ApJ...891..113R} found that the velocity dispersion of the molecular gas in this region shows larger values (mean $\sim 0.37$ \kms) compared to the values derived in the Southern Pillars (mean $\sim 0.31$ \kms), providing evidence for a higher level of turbulence in the shock/ionization fronts present in Car \ion{}{1}.  In fact, shock tracers such as SiO are easily detected along the fronts as is shown in Figure \ref{NC_images_velo}.  As the ionization fronts advance through the molecular gas in the Northern Cloud, they compress the low density gas, and move around regions of denser material, producing the arc-like structures identified in the continuum images and in the molecular gas map.  Inside these shocked regions the gas becomes unstable, collapsing into dense cores as observed by our high resolution ALMA maps.  As the erosion produced by the young stellar population continues and grows, the gas will be pushed backwards by the feedback, producing structure similar to the objects observed in the Southern Pillars, and ultimately, as the cometary object discussed in Section \ref{cometary}.

\section{SUMMARY}\label{summary}

In this paper we have presented the most detailed radio continuum map at $1-3$  GHz towards the CNC-Gum 31 molecular complex.  The observing program with the Australia Telescope Compact Array included 11 array configurations to provide a uniformly sampled  $u$-$v$ plane.  We covered $\sim$12 $\deg^2$, centered at $l = 287.5\degrees, b = -1\degrees$, achieving an angular resolution of $\sim 23\arcsec$.  We summarize our main results as follows:

\begin{enumerate}

\item The continuum map shows a spectacular and complex distribution of emission across the nebula, including structures such as filaments, shells, and arcs of widely different size scales.  The emission is highly concentrated in the center of the nuebula, powered by the massive star clusters Trumpler 14 and 16.  

\item We identified ionization fronts at distances $\sim$ 80 pc from the stellar clusters in the center towards the south of the Galactic Plane, and $\sim 50$ pc to the north.  Also in the north, a large shell-like structure of radius $\sim$ 10 pc was identified, with no stars detected inside. This structure is likely to have been formed by the pressure from hot plasma from the main \hii\ region.  

\item We estimated an ionization photon luminosity $Q_\mathrm{H}=(7.8 \pm 0.8) \times 10^{50}$ s$^{-1}$, which is $(15 \pm 8)\%$ less than the total value obtained from stellar population studies.  This ionization flux might have escaped from the nebula through multiples cavities and holes that are carved in the molecular cloud.

\item Comparison between the radio continuum and the \HI\ 21 cm map provided a better view of the bipolar outflow driven by the energy released by the massive stellar clusters.  Absorption features obtained against the bright continuum emission allowed us to detect the velocity components of the cold atomic gas.  Two main components are clearly detected, and they might related to the velocity of the expanding shells of ionized gas.

\item The CO maps revealed that individual clouds in Carina suffer different levels of the radiation, modifying the fraction of molecular gas.  The non-uniform advance of the ionized gas towards different regions in Carina might be related to different gas densities present in the parent cloud that gave birth to the first generation of massive stars.

\item Spatial anti-correlation between the 8 $\mu$m map and the radio continuum is observed in some regions in the CNC-Gum 31 complex.  Our high resolution map provides a detailed view of the permeability of the molecular cloud to the hot gas, which shapes the surrounding material into features such as pillars, small shells and arc-like structures, before ultimately escaping into the ISM.  

\item Some spatial correlation between warm dust as traced by 70 $\mu$m emission and our radio continuum image is detected.  This correlation is interpreted as the result of {\it in situ} heating of dust particles that are intermixed with the ionized gas in several regions in the nebula, particularly in the gas surrounding the surfaces of the clouds.

\item Two cometary objects are clearly identified in the radio continuum image of the center of Carina.  These structures appear to point to the Collinder 228 and Trumpler 16 clusters.  We suggest that these structures are in an advance stage of evolution of triggered star formation process, with little gas remaining due to the continuous erosion from the nearby stars after multiple episodes of star formation.

\item Our radio continuum image also revealed an object similar to a bipolar jet or outflow.  We suspect that this object might be a radio galaxy or a quasar not previously identified.  The emission morphology of the source shows bright spots at the very end of each opposite lobes, typical of extragalactic radio sources cataloged as FR Class II objects.

\item Unprecedented high spatial resolution image of the $1-3$ GHz radio continuum towards Car \ion{}{1} revealed multiple features such as filaments and arc-like structures.  These features are related to multiple ionization fronts previously detected at other radio frequencies.  Some diffuse radio continuum emission detected with ALMA at 3 mm shows some spatial correlation with the $1-3$ radio continuum map, while some other emission is more correlated with the molecular gas.  This provides further evidence for {\it in situ} heating of dust particles in \hii\ regions such as Car \ion{}{1}.

\end{enumerate}

\section*{Acknowledgements}
This paper makes use of the following ALMA data: ADS/JAO.ALMA\#2016.1.01609.S. ALMA is a partnership of ESO (representing its member states), NSF (USA) and NINS (Japan), together with NRC (Canada), MOST and ASIAA (Taiwan), and KASI (Republic of Korea), in cooperation with the Republic of Chile. The Joint ALMA Observatory is operated by ESO, AUI/NRAO and NAOJ. The National Radio Astronomy Observatory is a facility of the National Science Foundation operated under cooperative agreement by Associated Universities, Inc.  The Australia Telescope Compact Array and the Mopra Telescope are part of the Australia Telescope National Facility, funded by the Australian Government  and managed by CSIRO. This paper has made use of publicly available data from the Molonglo Observatory Synthesis Telescope Galactic Plane Survey at 843 MHz, the Herschel Galactic Plane Survey at 70 micron and the Spitzer IRAC survey at 8 micron.  DR acknowledges support from the ARC Discovery Project Grant DP130100338, and from CONICYT through project PFB-06 and project Fondecyt 3170568. GG acknowledges support from ANID project Basal AFB-170002.

\appendix

\section{RFI REMOVAL}\label{app_rfi}
\begin{figure*}[tbph]
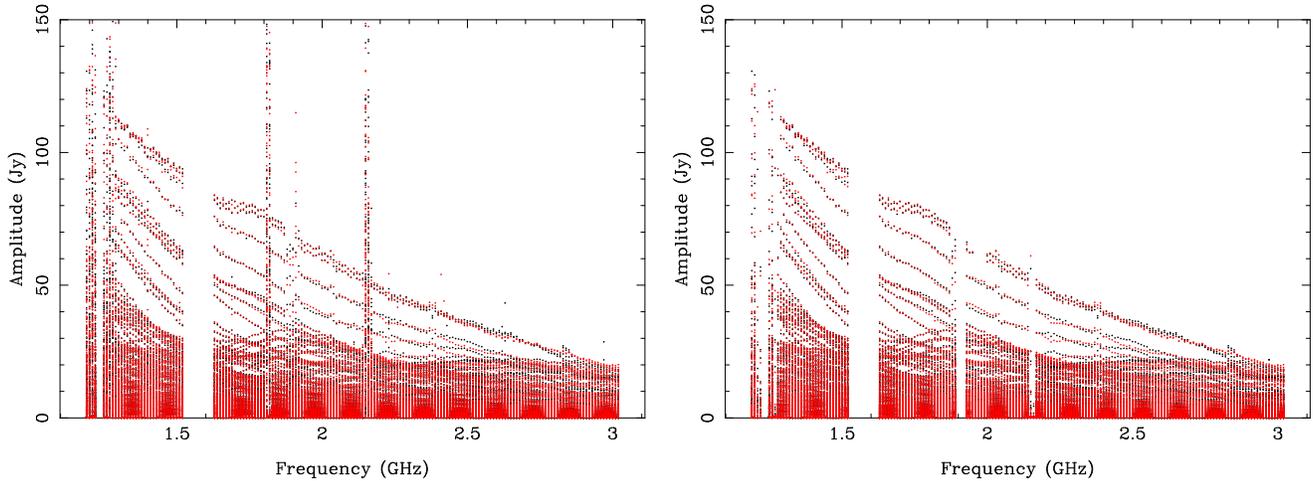

\centering
\begin{tabular}{cc}
\epsfig{file=Figure16_a.ps,width=0.35\linewidth,angle=-90} & \epsfig{file=Figure16_b.ps,width=0.35\linewidth,angle=-90} 
\end{tabular}
\caption{Example of the RFI in the spectrum towards one of fields (331) covered by our observations.  Different colors show the different polarization XX (black) and YY (red), including all the ATCA baselines. Because this field is located at the center of the nebula the emission shows complex structure at all baselines.  Left panel shows the spectrum without flagging, while the right panel shows the resulting clean spectrum after automatic and manual flagging has been applied.  Given the complex emission structure of the source across the spectrum, visual inspection of the spectra at different mosaic fields and baselines was needed in order to successfully remove the channels affected by the RFI events.}
\label{rfi_figure}
\end{figure*}

Radio Frequency Interference (RFI) is strong in the frequency range covered by our observations.  RFI is typically a transient phenomenon, and they can affect interferometry observations at different frequencies.  In addition, antenna baselines are affected differently, making removal more difficult.  For this project, short baselines are more affected for RFI events than the long ones, introducing another difficulty in the imaging process, particularly regarding the recovery of the diffuse emission.  The RFI removal strategy is complex and any approach to remove contaminated visibilities can be automatized to some extent.  Beyond that level, any residual RFI has to be removed manually, with a careful visual inspection of the visibilities versus time, frequency, baseline and pointing center.  Figure \ref{rfi_figure} shows the spectrum of a selected field located at the center of the nebula.  Channels affected by strong RFIs are evident as they are saturated in amplitude measurements with extremely large numbers.  Fortunately, in most of the cases, interference only affects a narrow range of channels, typically covering between 20 MHz and 40 MHz.  One exception is the large RFI region of $\sim$ 100 MHz extent located at $\sim$ 1.6 GHz.  

All the contaminated channels were flagged using the MIRIAD task PGFLAG.  We ran PGFLAG in automatic mode, using the SumThreshold method (\citealt{2010MNRAS.405..155O}) with a given number of parameters to guide the flagging process.  SumThreshold method performs a surface fit in the time-frequency plane in order to differentiate the astronomical signal from the RFI.  Several parameters used by SumThreshold can be fine-tuned to maximize the detection of the RFIs, including the amplitude threshold, convolution size in channel and time direction, and the level of propagation of the flags towards neighboring samples.  We found that using threshold values from 30 to 40 sigmas, and extending the flags to adjacent visibilities with less than 20\% of good data produces good results.  Although PGFLAG is efficient at removing RFIs, the method is not perfect and manual intervention was needed to remove residual RFI features in the spectra.  This second iteration of RFI cleaning was performed following a baseline by baseline, and pointing by pointing approach, which was the most time-consuming task of the calibration process.

\bibliography{biblio}

\clearpage


\end{document}